\documentclass[aip,reprint,superscriptaddress,longbibliography,floatfix]{revtex4-1}

\usepackage{amsmath,braket}
\usepackage{graphicx}
\usepackage{mathtools}
\usepackage{color}
\usepackage{xcolor}
\usepackage{setspace}

\makeatletter
\let\@fnsymbol\@fnsymbol@latex
\@booleanfalse\altaffilletter@sw
\makeatother

\begin{document}
\title{A gated quantum dot far in the strong-coupling regime of cavity-QED at optical frequencies}

\author{Daniel Najer}
\email{daniel.najer@unibas.ch}
\affiliation{Department of Physics, University of Basel, Klingelbergstrasse 82, CH-4056 Basel, Switzerland}

\author{Immo S\"{o}llner}
\affiliation{Department of Physics, University of Basel, Klingelbergstrasse 82, CH-4056 Basel, Switzerland}

\author{Pavel Sekatski}
\affiliation{Department of Physics, University of Basel, Klingelbergstrasse 82, CH-4056 Basel, Switzerland}

\author{Vincent Dolique}
\affiliation{Laboratoire des Mat\'{e}riaux Avanc\'{e}s (LMA), IN2P3/CNRS, Universit\'{e} de Lyon, F-69622 Villeurbanne, Lyon, France}

\author{Matthias C.\ L\"{o}bl}
\affiliation{Department of Physics, University of Basel, Klingelbergstrasse 82, CH-4056 Basel, Switzerland}

\author{Daniel Riedel}
\affiliation{Department of Physics, University of Basel, Klingelbergstrasse 82, CH-4056 Basel, Switzerland}

\author{R\"{u}diger Schott}
\affiliation{Lehrstuhl f\"{u}r Angewandte Festk\"{o}rperphysik, Ruhr-Universit\"{a}t Bochum, D-44780 Bochum, Germany}

\author{Sebastian Starosielec}
\affiliation{Department of Physics, University of Basel, Klingelbergstrasse 82, CH-4056 Basel, Switzerland}

\author{Sascha R.\ Valentin}
\affiliation{Lehrstuhl f\"{u}r Angewandte Festk\"{o}rperphysik, Ruhr-Universit\"{a}t Bochum, D-44780 Bochum, Germany}

\author{Andreas D.\ Wieck}
\affiliation{Lehrstuhl f\"{u}r Angewandte Festk\"{o}rperphysik, Ruhr-Universit\"{a}t Bochum, D-44780 Bochum, Germany}

\author{Nicolas Sangouard}
\affiliation{Department of Physics, University of Basel, Klingelbergstrasse 82, CH-4056 Basel, Switzerland}

\author{Arne Ludwig}
\affiliation{Lehrstuhl f\"{u}r Angewandte Festk\"{o}rperphysik, Ruhr-Universit\"{a}t Bochum, D-44780 Bochum, Germany}

\author{Richard J.\ Warburton}
\affiliation{Department of Physics, University of Basel, Klingelbergstrasse 82, CH-4056 Basel, Switzerland}

\date{\today}

\begin{abstract}
The strong-coupling regime of cavity-quantum-electrodynamics (cQED) represents light-matter interaction at the fully quantum level. Adding a single photon shifts the resonance frequencies, a profound nonlinearity. cQED is a test-bed of quantum optics~\cite{Boca2004,Birnbaum2005,Hamsen2017} and the basis of photon-photon and atom-atom entangling gates~\cite{Zheng2000,Duan2004}. At microwave frequencies, success in cQED has had a transformative effect~\cite{Fink2008}. At optical frequencies, the gates are potentially much faster and the photons can propagate over long distances and be easily detected, ideal features for quantum networks. Following pioneering work on single atoms~\cite{Boca2004,Birnbaum2005,Hamsen2017,Vuletic2018}, solid-state implementations are important for developing practicable quantum technology~\cite{Reithmaier2004,Yoshie2004,Faraon2008,Hennessy2007,Rakher2009,Reinhard2012,Volz2012,Ota2018}. Here, we embed a semiconductor quantum dot in a microcavity. The microcavity has a $\mathcal{Q}$-factor close to $10^{6}$ and contains a charge-tunable quantum dot with close-to-transform-limited optical linewidth. The exciton-photon coupling rate $g$ exceeds both the photon decay rate $\kappa$ and exciton decay rate $\gamma$ by a large margin ($g/\gamma=14$, $g/\kappa=5.3$); the cooperativity is $C=2g^{2}/(\gamma \kappa)=150$, the $\beta$-factor 99.7\%. We observe pronounced vacuum Rabi oscillations in the time-domain, photon blockade at a one-photon resonance, and highly bunched photon statistics at a two-photon resonance. We use the change in photon statistics as a sensitive spectral probe of transitions between the first and second rungs of the Jaynes-Cummings ladder. All experiments can be described quantitatively with the Jaynes-Cummings model despite the complexity of the solid-state environment. We propose this system as a platform to develop optical-cQED for quantum technology, for instance a photon-photon entangling gate.
\end{abstract}

\maketitle
\noindent An excellent solid-state emitter of single photons is a self-assembled quantum dot in a semiconductor host~\cite{Kuhlmann2013,Somaschi2016}. An InGaAs semiconductor quantum dot in GaAs is a bright and fast emitter of highly indistinguishable photons, properties not shared by any other emitter. The challenge in pursuing the strong-coupling regime of cQED with such a quantum dot is to combine apparently contradictory elements.

First, the cavity must have an ultrahigh $\mathcal{Q}$-factor yet a small mode volume, i.e.\ dimensions comparable to the optical wavelength. Nano-fabrication techniques are employed to create, for instance, micropillar~\cite{Reithmaier2004,Somaschi2016} or photonic crystal cavities~\cite{Yoshie2004,Faraon2008,Hennessy2007,Reinhard2012,Volz2012,Ota2018}. The acute problem is that the $\mathcal{Q}$-factor tends to deteriorate as the mode volume decreases. This is only partly a consequence of fabrication imperfections, sidewall roughness of a micropillar for example. An additional factor is the GaAs surface which pins the Fermi energy mid-gap resulting in surface-related absorption~\cite{Guha2017}. Achieving a low-volume, ultrahigh $\mathcal{Q}$-factor cavity in GaAs has proved to be difficult. Secondly, a quantum dot benefits enormously from electrical control via the conducting gates of a diode structure. A gated quantum dot in high quality material gives close-to-transform-limited linewidths~\cite{Kuhlmann2013} and control over both the optical frequency via the Stark effect and the quantum dot charge state~\cite{Hoegele2004}. A charge-neutral quantum dot operates as a two-level system and is ideal as source of highly indistinguishable photons; a single electron or hole allows the creation of entangled spin-photon pairs. However, the conducting layers of gated devices are not obviously compatible with an ultrahigh ${\mathcal Q}$-factor cavity on account of significant free-carrier absorption in the doped layers and below-band-gap absorption via the Franz-Keldysh effect. Finally, the quantum dot in a microcavity must retain the close-to-transform-limited optical linewidths of the starting material. This is hard to achieve following aggressive nano-fabrication as the free surface can result in additional charge noise leading to blinking and spectral fluctuations of the quantum dot.

\begin{figure*}[t!]
\centering
\includegraphics[width=\textwidth]{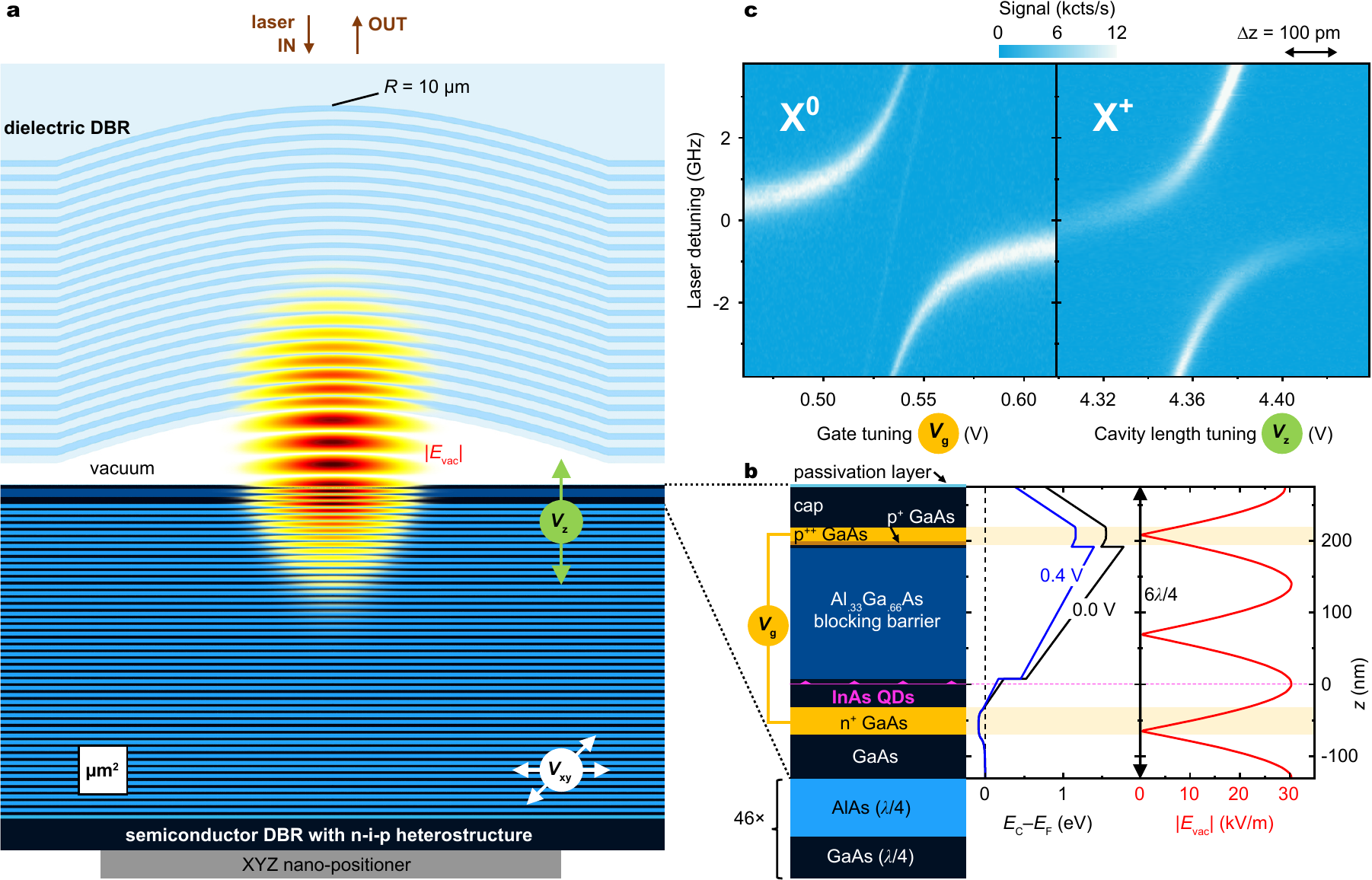}

\caption{{\bf Gated quantum dot in a tunable microcavity: design and realisation.} {\bf a}, Simulation of the vacuum electric field $|E_{\rm vac}|$ in the microcavity (image to scale). The bottom mirror is a distributed Bragg reflector (DBR) consisting of 46 AlAs($\lambda/4$)/GaAs($\lambda/4$) pairs. ($\lambda$ refers to the wavelength in each material.) The top mirror is fabricated in a silica substrate \cite{Barbour2011,Hunger2012}. It has radius of curvature $R=10$\,$\mu$m and consists of 22 silica($\lambda/4$)/tantala($\lambda/4$) pairs. The layer of quantum dots (QDs) is located at the vacuum field anti-node one wavelength beneath the surface. The vacuum-gap has the dimension of $3\lambda/2$. $V_{\rm xy}$ ($V_{\rm z}$) controls the lateral (vertical) position of the QD with respect to the fixed top mirror. {\bf b}, The top part of the semiconductor heterostructure. A voltage $V_{\rm g}$ is applied across the n-i-p diode. $V_{\rm g}$ controls the QD-charge via Coulomb blockade and within a Coulomb blockade plateau the exact QD optical frequency via the dc Stark effect. Free-carrier absorption in the p-layer~\cite{Casey1975} is minimised by positioning it at a node of the vacuum field. A passivation layer suppresses surface-related absorption~\cite{Guha2017}. {\bf c}, Laser detuning ($\Delta_{\rm L}$) versus cavity detuning ($\Delta_{\rm C}$) of a neutral QD exciton (X$^{0}$) and a positively-charged exciton (X$^{+}$) in one and the same QD (QD1). Cavity detuning is achieved by tuning the QD at fixed microcavity frequency (X$^{0}$); and by tuning the microcavity frequency at fixed QD frequency (X$^{+}$). For X$^{0}$, the weak signal close to the bare microcavity frequency arises from weak coupling to the other orthogonally-polarised X$^{0}$ transition -- it does not arise from blinking (see Supplementary III.E).}

\label{design}
\end{figure*}

We present a resolution to these conundrums. We have found a way to create an ultrahigh $\mathcal{Q}$-factor yet with small mode volume. The quantum dot is gated and exhibits close-to-transform-limited optical linewidths even in the cavity. On resonance with the microcavity, the quantum dot exciton is far in the strong-coupling regime. Strong coupling is achieved on both neutral and charged excitons in one and the same quantum dot by tuning the microcavity {\em in situ}. The output is close to a simple Gaussian beam allowing high efficiency collection. Notably, the solid-state feature which has complicated quantum dot cQED in the past -- scattering from the bare cavity mode even at the quantum dot-cavity resonance~\cite{Hennessy2007,Faraon2008, Reinhard2012, Volz2012, Ota2018, KurumaRRB2018, Greuter2015} -- disappears. The system is an exemplary Jaynes-Cummings system despite the complexity of the solid-state environment. 


\begin{figure*}[t!]
\centering
\includegraphics[width=\textwidth]{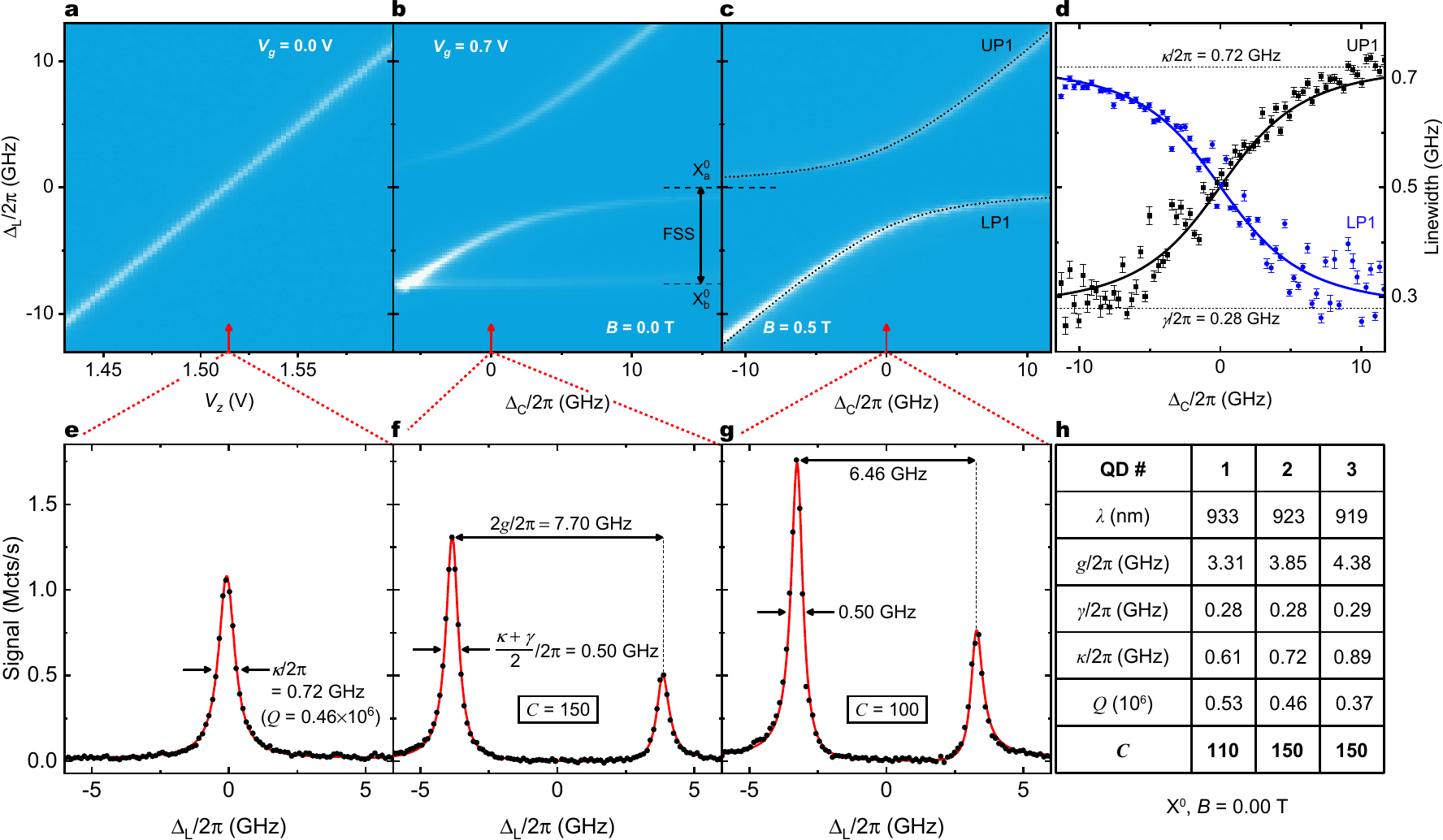}

\caption{{\bf Strong coupling of a QD exciton in the microcavity.} The spectra were recorded by measuring the photons scattered by the microcavity--QD system at a temperature of 4.2\,K, rejecting reflected laser light with a polarisation-based dark-field technique~\cite{Kuhlmann2013}. Data shown here were taken on the X$^{0}$ transition (QD2). {\bf a}, {\bf e} signal with QD far-detuned from microcavity in order to determine the photon loss-rate $\kappa$, equivalently the quality factor $\mathcal{Q}$). {\bf b}, {\bf f} X$^{0}$ at magnetic field $B=0.00$\,T showing strong coupling to one fine-structure-split (FSS) transition, weak coupling to the other (there is an almost perfect alignment of the X$^{0}$ and microcavity axes). From the spectra, we determine the X$^{0}$--vacuum-field coupling rate ($g$) and the QD exciton decay rate into other photonic modes ($\gamma$). The cooperativity is defined as $C=2g^2/(\kappa \gamma)$. {\bf c}, {\bf d}, {\bf g}, X$^{0}$ at $B=0.50$\,T: the magnetic field induces a large frequency separation between the fine-structure-split transitions. $C$ is smaller than at $B=0$ because the X$^{0}$ transitions become circularly polarised and couple less strongly to the linear-polarised microcavity mode. The simple avoided-crossing in {\bf c} enables a determination of $\kappa$ and $\gamma$ by using data at all values of $\Delta_{\rm C}$. The dotted lines in {\bf c} and solid lines in {\bf d}--{\bf g} are fits to a solution of the Jaynes-Cummings Hamiltonian in the limit of very small average photon occupation~\cite{Greuter2015}. {\bf h}, Summary of strong-coupling parameters recorded on X$^{0}$ at $B=0.00$\,T on three separate QDs using the same microcavity mode. $C>100$ in all three cases.}

\label{spectroscopy}
\end{figure*}

We employ a miniaturised Fabry-P\'{e}rot cavity consisting of a semiconductor heterostructure and external top mirror (Fig.\ \ref{design}a and Supplementary section II). The $\mathcal{Q}$-factor is as high as $10^{6}$; the mode volume just $1.4 \lambda_0^{3}$ (where $\lambda_{\rm 0}$ is the free-space wavelength). The heterostructure (see Supplementary section I) has an n-i-p design with the quantum dots in the intrinsic (i) region (Fig.\ \ref{design}b). Tunnel contact with the Fermi sea in the n-type layer establishes charge control via Coulomb blockade. 

We excite the quantum dot--microcavity system with a resonant laser (continuous-wave) and detect the scattered photons. The average photon occupation is much less than one.
When the microcavity and QD optical frequency come into resonance, we observe a clear avoided crossing in the spectral response (Fig.\ \ref{design}c) signifying strong coupling. We achieve strong coupling on different charge states in the same QD (Fig.\ \ref{design}c), also on many different QDs (Fig.\ \ref{spectroscopy}h and Supplementary section III). The cavity-emitter detuning 
is controlled {\em in situ} either by tuning the QD (voltage $V_{\rm g}$) or by tuning the microcavity (voltage $V_{\rm z}$) (Fig.\ \ref{design}c). A full spectral analysis determines the exciton-photon coupling rate $g$, the cavity-photon decay rate $\kappa$, and the exciton decay rate into non-microcavity modes, $\gamma$ (Fig.\ \ref{spectroscopy}). For QD2 at zero magnetic field, $g/\gamma=14$, $g/\kappa=5.3$ corresponding to a cooperativity $C=2 g^{2}/(\kappa \gamma)=150$. 
The $\beta$-factor~\cite{Kuhn2010}, the fraction of quantum dot emission funnelled into the cavity mode, is $\beta=2C/(2C+1)=99.7$\%.

\begin{figure}[t!]
\centering
\includegraphics[width=86 mm]{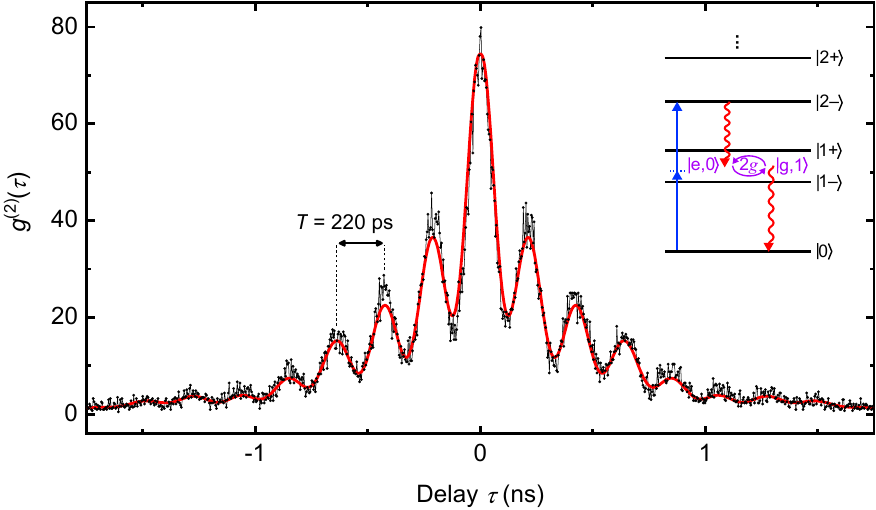}

\caption{{\bf Time-resolved vacuum Rabi oscillations.} Intensity auto-correlation function $g^{(2)}(\tau)$ as a function of delay $\tau$ on X$^{0}$ in QD1 for $\Delta_{\rm C}=0.73 g$ (detuned via $V_z$) and $\Delta_{\rm L}=-0.13 g$. The inset shows the first few rungs of the Jaynes-Cummings ladder. The laser drives a two-photon transition $\ket{0} \leftrightarrow \ket{2-}$. The solid red line is the result of calculating $g^{(2)}(\tau)$ from the Jaynes-Cummings Hamiltonian using $g$, $\kappa$ and $\gamma$ from the spectroscopy experiments (Fig.\ \ref{spectroscopy}).}

\label{g2}
\end{figure}

\begin{figure*}[t!]
\centering
\includegraphics[width=\textwidth]{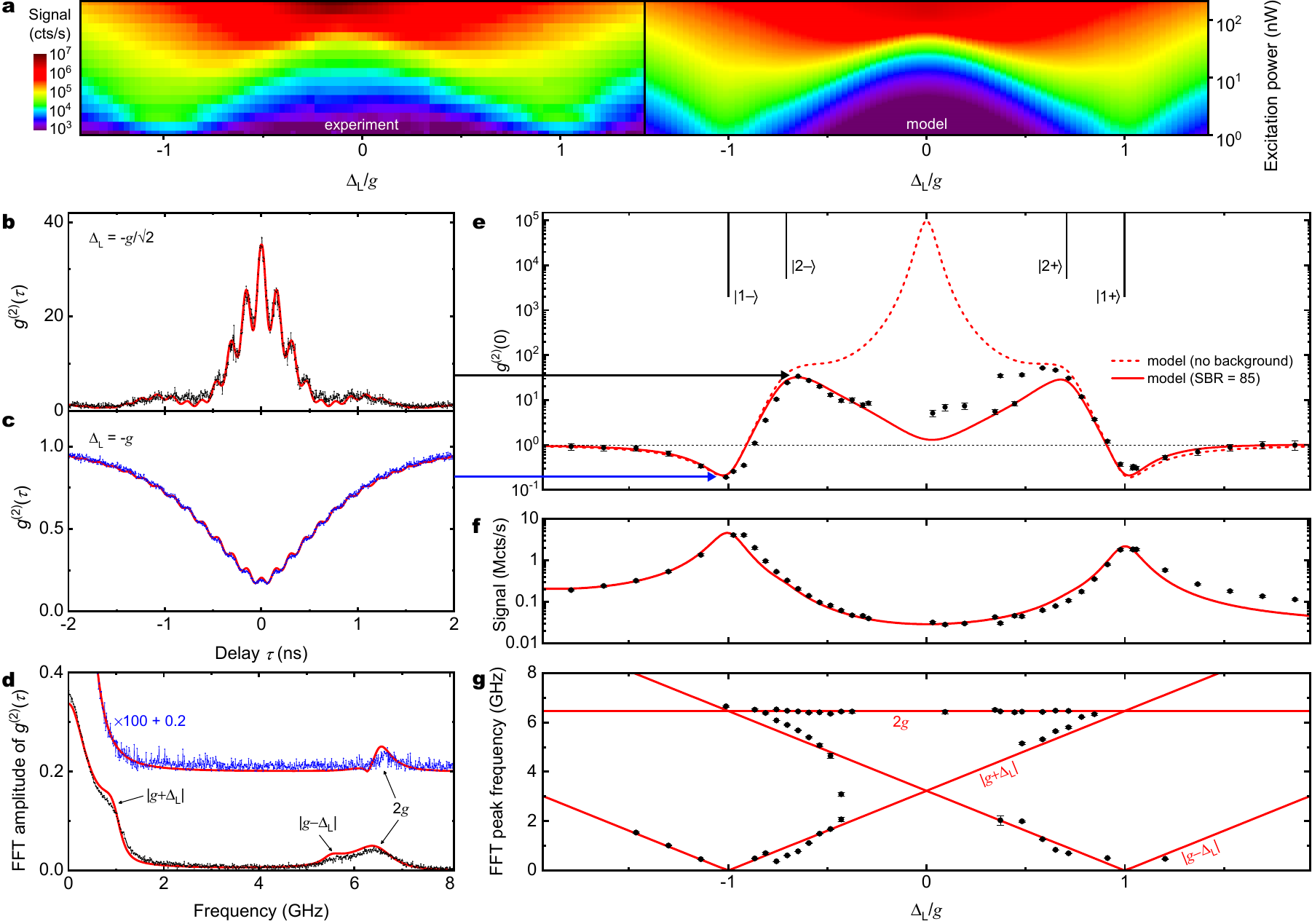}

\caption{{\bf Strong coupling versus driving frequency and power.} {\bf a}, Signal versus $\Delta_{\rm L}$ for $\Delta_{\rm C}=0$. At low power, LP1 and UP1 are clearly observed. As the power increases, the higher rungs of the Jaynes-Cummings ladder are populated. {\bf b}, $g^{(2)}(\tau)$ for $\Delta_{\rm C}=0$ and $\Delta_{\rm L}=-g/\sqrt{2}$. {\bf c}, $g^{(2)}(\tau)$ for $\Delta_{\rm C}=0$ and $\Delta_{\rm L}=-g$. {\bf d}, fast Fourier transform (FFT) of $g^{(2)}(\tau)$ in {\bf b} and {\bf c}. {\bf e}, {\bf f} and {\bf g}: $g^{(2)}(0)$, signal and FFT peak frequency of $g^{(2)}(\tau)$ versus $\Delta_{\rm L}$ for $\Delta_{\rm C}=0$. The solid red lines in {\bf b}--{\bf g} (``model" in {\bf a}) result from a calculation of $g^{(2)}(\tau)$ (signal) from the Jaynes-Cummings Hamiltonian using $g$, $\kappa$ and $\gamma$ from the spectroscopy experiments (Fig.\ \ref{spectroscopy}). In {\bf a}, the truncation of the Hilbert space to 15 rungs leads to a slight underestimation of the signal at high laser powers compared to the experiment. A signal-to-background ratio (SBR) of 85 was included. In {\bf e}, the dashed red line shows the theoretical limit without the laser background.}

\label{g2FT}
\end{figure*}

To demonstrate a coherent atom-photon exchange, ``vacuum Rabi oscillations''~\cite{KasprzakNatMat2010,Hamsen2017,KurumaRRB2018}, we drive the system at a frequency slightly positively-detuned from the lower-frequency polariton (LP1) and record the two-photon auto-correlation $g^{(2)}(\tau)$ (Fig.\ \ref{g2}). Coherent oscillations are observed as a function of delay whose period, 220 ps, corresponds exactly to $2\pi$ divided by the measured frequency splitting of the polaritons (Supplementary section III.F).
These oscillations can be understood in terms of the Jaynes-Cummings ladder (Fig.\ \ref{g2} inset). The laser drives the two-photon transition $\ket{0} \leftrightarrow \ket{2-}$ weakly.
$\ket{2-}$ decays by emitting two photons. Detection of the first photon leaves the system in a superposition of the eigenstates $\ket{1-}$ and $\ket{1+}$ such that a quantum beat takes place. Detection of the second photon projects the system into the ground state $\ket{0}$, stopping the quantum beat (Supplementary section V).
The large $g^{(2)}(0)$ (80 in this particular experiment) is confirmation that the states with $n\geq 2$ are preferentially scattered~\cite{Faraon2008,Reinhard2012}. 

The behaviour of $g^{(2)}(\tau)$ depends strongly on the laser detuning $\Delta_{\rm L}$ and the cavity detuning $\Delta_{\rm C}$ (both defined with respect to the bare exciton). For $\Delta_{\rm C}= 0$, $g^{(2)}(0)$ is highly bunched at the two-photon resonance, $\Delta_{\rm L}=-g/\sqrt{2}$ (Fig.\ \ref{g2FT}b), yet highly anti-bunched at the single-photon resonance, $\Delta_{\rm L}=-g$ (Fig.\ \ref{g2FT}c). The anti-bunching is a demonstration of photon blockade in this system. The full dependence on $\Delta_{\rm L}$ is plotted in Fig.\ \ref{g2FT}e. In principle, $g^{(2)}(0)$ rises to extremely high values~\cite{Birnbaum2005} as $\Delta_{\rm L} \rightarrow 0$. In practice, the scattered signal becomes weaker and weaker as $\Delta_{\rm L} \rightarrow 0$ such that $g^{(2)}(0)$ reaches a peak and is then pulled down by the poissonian statistics of the small leakage of laser light into the detector channel (Fig.\ \ref{g2FT}e). $g^{(2)}(\tau)$ is a rich function: its Fourier transform shows in general three peaks (Fig.\ \ref{g2FT}d). The dependence on $\Delta_{\rm L}$ shows that these frequencies correspond to $2g$ (see Supplementary section V.D.3), $|g-\Delta_{\rm L}|$ and $|g+\Delta_{\rm L}|$ (Fig.\ \ref{g2FT}g).
All this complexity is described by the Jaynes-Cummings model which, taking the parameters determined by the spectroscopy experiments and a numerical solution using the first fifteen rungs of the ladder (Supplementary section IV), gives excellent agreement with the experimental $g^{(2)}(\tau)$ in all respects (Fig.\ \ref{g2}, Fig.\ \ref{g2FT} and Supplementary section III). As the laser power increases, there is a spectral resonance at the LP2 and UP2 transitions, and, at the highest powers, a strong resonance at $\Delta_{\rm L}=0$ -- this too is in agreement with the predictions of the model (Fig.\ \ref{g2FT}a), and reflects the bosonic enhancement of the transitions between the higher lying rungs of the Jaynes-Cummings ladder.

\begin{figure}[t!]
\centering
\includegraphics[width=86 mm]{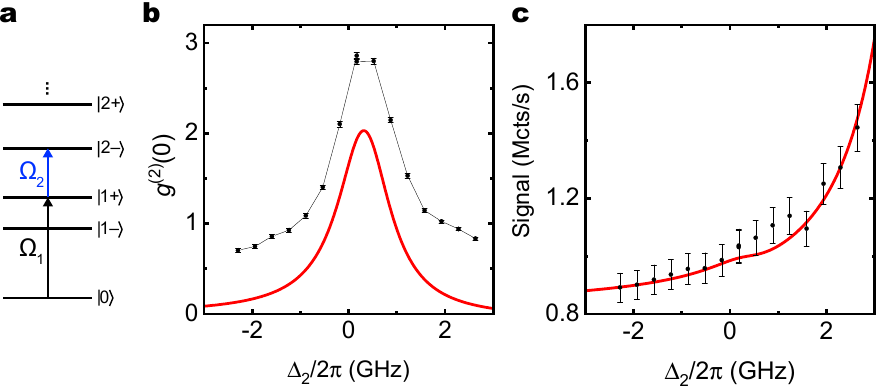}

\caption{{\boldmath $g^{(2)}$}{\bf -spectroscopy.} {\bf a}, Laser 1 is on resonance with the $\ket{0} \leftrightarrow \ket{1+}$ transition (black arrow, detuning $\Delta_{1}=0$); laser 2 is scanned across the $\ket{1+} \leftrightarrow \ket{2-}$ transition (blue arrow, detuning $\Delta_{2}$). {\bf b}, $g^{(2)}(0)$ versus $\Delta_{2}$ showing a pronounced resonance at $\Delta_{2}=3\Delta_{\rm C}/2-\Delta_{1}$. The red solid line is the result of an analytical calculation based on the Jaynes-Cummings Hamiltonian (Supplementary section VI). The offset in the experimental data with respect to the theory reflects additional coincidences arising from off-resonant, two-photon absorptions not included in the model. {\bf c}, Signal versus $\Delta_{2}$. The signal increases with increasing $\Delta_{2}$ due to off-resonant driving of the $\ket{0} \leftrightarrow \ket{1-}$ transition by laser 2. All data for X$^{0}$ in QD2 at $B=0.50$\,T.}

\label{g20}
\end{figure}

In the experiments with a single laser, the second rung of the Jaynes-Cummings ladder is accessed by tuning the laser to a two-photon resonance (Fig.\ \ref{g2FT}a). An alternative is to drive the system with two lasers in a pump-probe scheme. The strong transitions arise from the symmetric-to-symmetric and antisymmetric-to-antisymmetric couplings, e.g.\ $\ket{1-} \leftrightarrow \ket{2-}$ and $\ket{1+} \leftrightarrow \ket{2+}$, which lead to measurable changes in the populations of the states~\cite{Fink2008}. We present an alternative here, ``$g^{(2)}$-spectroscopy". We present this experiment on the symmetric-to-asymmetric $\ket{1+} \leftrightarrow \ket{2-}$ transition. The square of the matrix element is just 3\% of that associated with the $\ket{1+} \leftrightarrow \ket{2+}$ transition. A pump laser drives the $\ket{0} \leftrightarrow \ket{1+}$ transition on resonance, and a probe laser, highly red-detuned with respect to the pump, is scanned in frequency in an attempt to locate the $\ket{1+} \leftrightarrow \ket{2-}$ transition (Fig.\ \ref{g20}a). There is no resonance in the scattered intensity (Fig.\ \ref{g20}c): any resonance lies in the noise (a few per cent). However, there is a clear resonance in $g^{(2)}(0)$ at exactly the expected frequency $\Delta_{2}=3\Delta_{\rm C}/2-\Delta_{1}$ (Fig.\ \ref{g20}b): at the weak $\ket{1+} \leftrightarrow \ket{2-}$ transition the number of scattered photons hardly changes but there are profound changes in their statistical correlations. Again, the Jaynes-Cummings model describes the experiment (Fig.\ \ref{g20}b,c). Here, a short-time expansion in a truncated Hilbert space (first two rungs of the Jaynes-Cummings ladder) is used to calculate $g^{(2)}(0)$ (Supplementary section VI). 

As an outlook, we offer some perspectives for future development. (a) The device is a potentially excellent single photon source. For a fixed $g$ and $\gamma$, the photon extraction efficiency via the cavity~\cite{CuiOptExp2005} is maximised at the condition $\kappa=2g$. For $g$ achieved here, this corresponds to $\mathcal{Q}=3.7\cdot 10^{4}$. At this relatively low $\mathcal{Q}$, the residual absorption losses in the semiconductor are negligible and the photon extraction efficiency should be as high as 90\%. 
(b) An ``atom drive"~\cite{Law1997,Hamsen2017} can be engineered with a lateral waveguide. This is an excellent prospect for creating fast spin-photon entanglements, shaped-waveform single photons and, ultimately, a photon-photon gate. (c) An even higher $C$ is conceivable by decreasing $\gamma$ via some lateral processing. (d) Two or more intra-cavity quantum dot spins can be entangled by common coupling to the cavity mode~\cite{Imamoglu1999}. (e) A monolithic design could exploit strain tuning of the quantum dot rather than position-based tuning of the cavity. Also, the splitting of the cavity mode (into two modes with linear, orthogonal polarisations) can be eliminated by applying a bias across the semiconductor DBR~\cite{Frey2018}.\\

\noindent{\bf Acknowledgements}\\
We thank Ivan Favero for inspiration on surface passivation; Henri Thyrrestrup Nielsen for support in evaluating $g^{(2)}(\tau)$ with very small binning times; Sascha Martin for engineering the microcavity hardware; and Melvyn Ho, Peter Lodahl and Philipp Treutlein for fruitful discussions. We acknowledge financial support from SNF projects 200020\_156637 and PP00P2\_179109, NCCR QSIT and EPPIC (747866). SRV, RS, AL and ADW acknowledge gratefully support from BMBF Q.com-H 16KIS0109.\\

\noindent{\bf Additional information}\\
Further details of the experiment and the calculations based on the Jaynes-Cummings Hamiltonian are described in the Supplementary Information.\\

\end{document}


\title{A gated quantum dot far in the strong-coupling regime of cavity-QED at optical frequencies\\ \vspace{10PT}
\small{Supplementary Information}}

\author{Daniel Najer}
\email{daniel.najer@unibas.ch}
\affiliation{Department of Physics, University of Basel, Klingelbergstrasse 82, CH-4056 Basel, Switzerland}

\author{Immo S\"{o}llner}
\affiliation{Department of Physics, University of Basel, Klingelbergstrasse 82, CH-4056 Basel, Switzerland}

\author{Pavel Sekatski}
\affiliation{Department of Physics, University of Basel, Klingelbergstrasse 82, CH-4056 Basel, Switzerland}

\author{Vincent Dolique}
\affiliation{Laboratoire des Mat\'{e}riaux Avanc\'{e}s (LMA), IN2P3/CNRS, Universit\'{e} de Lyon, F-69622 Villeurbanne, Lyon, France}

\author{Matthias C.\ L\"{o}bl}
\affiliation{Department of Physics, University of Basel, Klingelbergstrasse 82, CH-4056 Basel, Switzerland}

\author{Daniel Riedel}
\affiliation{Department of Physics, University of Basel, Klingelbergstrasse 82, CH-4056 Basel, Switzerland}

\author{R\"{u}diger Schott}
\affiliation{Lehrstuhl f\"{u}r Angewandte Festk\"{o}rperphysik, Ruhr-Universit\"{a}t Bochum, D-44780 Bochum, Germany}

\author{Sebastian Starosielec}
\affiliation{Department of Physics, University of Basel, Klingelbergstrasse 82, CH-4056 Basel, Switzerland}

\author{Sascha R.\ Valentin}
\affiliation{Lehrstuhl f\"{u}r Angewandte Festk\"{o}rperphysik, Ruhr-Universit\"{a}t Bochum, D-44780 Bochum, Germany}

\author{Andreas D.\ Wieck}
\affiliation{Lehrstuhl f\"{u}r Angewandte Festk\"{o}rperphysik, Ruhr-Universit\"{a}t Bochum, D-44780 Bochum, Germany}

\author{Nicolas Sangouard}
\affiliation{Department of Physics, University of Basel, Klingelbergstrasse 82, CH-4056 Basel, Switzerland}

\author{Arne Ludwig}
\affiliation{Lehrstuhl f\"{u}r Angewandte Festk\"{o}rperphysik, Ruhr-Universit\"{a}t Bochum, D-44780 Bochum, Germany}

\author{Richard J.\ Warburton}
\affiliation{Department of Physics, University of Basel, Klingelbergstrasse 82, CH-4056 Basel, Switzerland}

\date{\today}

\pacs{}

\maketitle 
\beginsupplement


\section{Semiconductor heterostructure}
\subsection{Design and growth}
The heterostructure is grown by molecular beam epitaxy (MBE). It consists of an n-i-p diode with embedded self-assembled InAs quantum dots grown on top of an AlAs/GaAs distributed Bragg reflector (DBR) with nominal (measured) centre wavelength of 940\,nm (920\,nm).

\begin{figure*}[t!]{}
	\centering
\includegraphics[width=\textwidth]{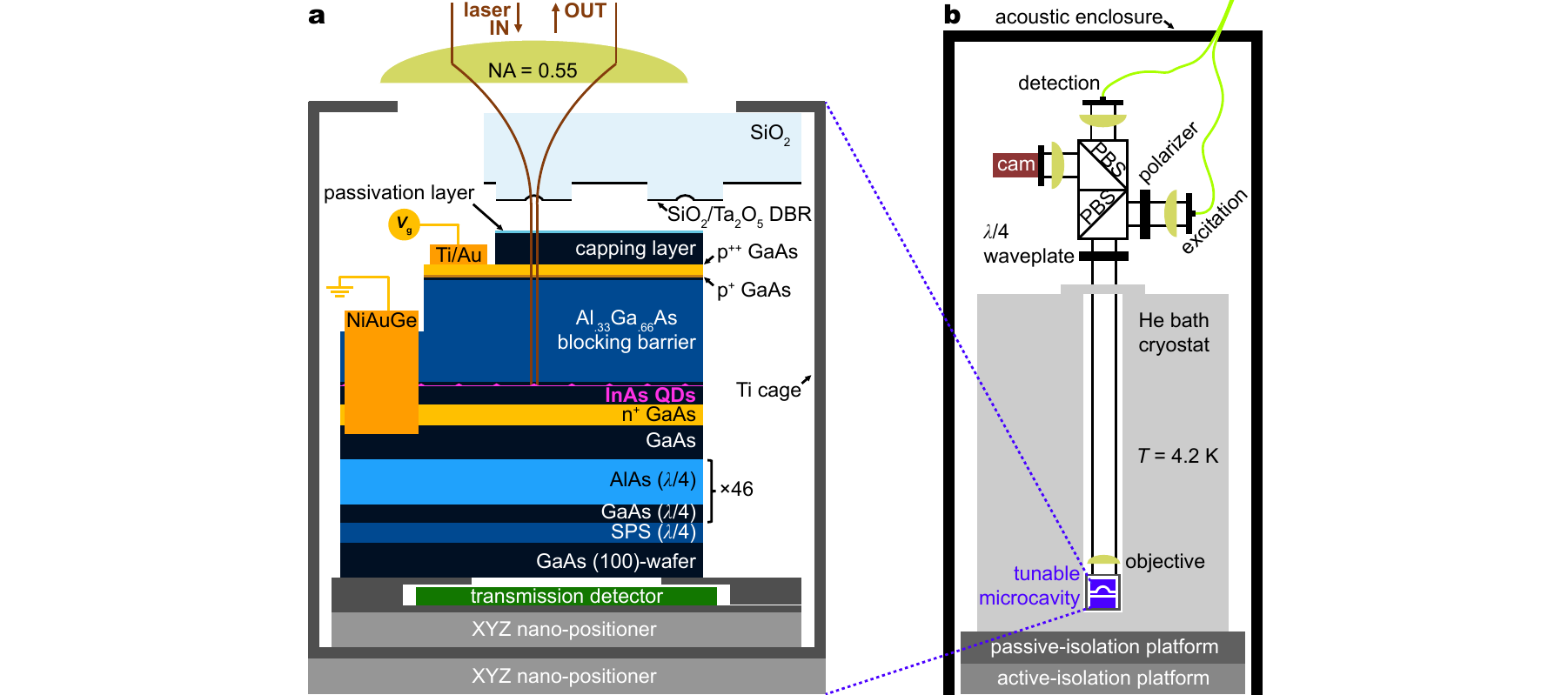}
\begin{singlespace}
 \caption{
 \textbf{Tunable microcavity setup.}
	\textbf{a} The top-mirror is fixed to the upper inner-surface of a titanium ``cage". The sample is mounted on a piezo-driven XYZ nano-positioner; the nano-positioner is fixed to the bottom inner-surface of the cage. The nano-positioner allows for full \textit{in situ} spatial and spectral tuning of the microcavity at cryogenic temperatures. The titanium cage resides on another XYZ nano-positioner allowing for close-to-perfect mode matching of the cavity mode to the external laser beam \cite{GreuterAPL2014}. \textbf{b} An outer Ti cage containing the inner Ti cage and second nano-positioner is fixed to an optical rod-system which is inserted into a vacuum tube filled with He exchange gas. The optical elements depicted in the image (objective lens, a quarter-wave plate, two polarising beam-splitters (PBSs), a polariser, a CMOS camera, two fibre couplers) make up the dark-field microscope for close-to-background-free detection of resonance fluorescence \cite{KuhlmannRSI2013}. The back-reflected laser is suppressed by a factor up to $10^8$ by choosing orthogonal polarisation states for the excitation and detection channels \cite{KuhlmannRSI2013}. The optical fibre attached to the microscope's excitation (detection) arm includes a 50:50 (99:1) fibre beam-splitter in order to monitor the laser power sent into the microscope (reflected from the sample). The cryostat sits on both active- and passive-isolation platforms and is surrounded by an acoustic enclosure to minimise acoustic noise. Both images are schematic representations and are not to scale. The exact layer thicknesses and doping concentrations are found in the text.
	}
	\label{fig:fig1_detailed}
\end{singlespace}
\end{figure*}

The growth on a (100)-oriented GaAs wafer is initiated by a quarter-wave layer (QWL) of an AlAs/GaAs short-period superlattice (SPS, 18 periods of 2.0\,nm GaAs and 2.0\,nm AlAs) for stress-relief and surface-smoothing. The growth continues with 46 pairs of GaAs (67.9\,nm) and AlAs (80.6\,nm) QWLs forming the ``bottom" DBR. The active part of the device consists of a QWL of GaAs (69.8\,nm) followed by a 41.0\,nm thick layer of Si-doped GaAs (n$^+$, $2\cdot10^{18}$\,cm$^{-3}$), the back-gate. 25.0\,nm of undoped GaAs, the tunnel barrier, is subsequently grown, after which InGaAs quantum dots are self-assembled using the Stranski-Krastanow process and a flushing-step~\cite{Wasilewski1999} to blue-shift the quantum dot emission. The layer thicknesses are such that the quantum dots are located at an antinode of the vacuum electric field. The quantum dots are capped with an 8.0\,nm layer of GaAs. The growth proceeds with an Al$_{.33}$Ga$_{.66}$As layer (190.4\,nm), a blocking barrier to reduce the current flow through the diode structure. The heterostructure is completed by 25.0\,nm C-doped GaAs (5.0\,nm p$^+$, $2\cdot10^{18}$\,cm$^{-3}$ and 20.0\,nm p$^{++}$, $1\cdot10^{19}$\,cm$^{-3}$), the top-gate, and, finally, a 54.6\,nm GaAs capping layer. The heterostructure is shown in Fig.~\ref{fig:fig1_detailed}. 

The top-gate is centred around a node of the standing wave of the vacuum electric field in order to minimise free-carrier absorption from the p-doped GaAs. A condition on the tunnel barrier thickness (it is typically $\lesssim 40$\,nm thick to achieve a non-negligible tunnel coupling with the Fermi sea) prevents the back-gate being positioned likewise at a vacuum field node. However, the free-carrier absorption of n$^+$-GaAs is much smaller than that of p$^{++}$-GaAs at a photon energy of 1.3\,eV ($\alpha\approx10$\,cm$^{-1}$ for n$^+$-GaAs compared to $\alpha\approx70$\,cm$^{-1}$ for p$^{++}$-GaAs \cite{CaseyJAP1975}). We exploit the weak free-carrier absorption of n$^+$-GaAs and use a standard 25\,nm thick tunnel barrier. The back-gate is thus positioned close to the node of the vacuum electric field but is not centred around the node itself.

\subsection{Post-growth processing: electrical contacts and surface passivation}
After growth, individual $2.5 \times 3.0$\,mm$^2$ pieces of the wafer are processed: separate ohmic contacts are made to the n$^{+}$ and p$^{++}$ layers; a passivation layer is added to the surface. To contact the n$^{+}$-layer, the back-gate, a local etch in citric acid is used to remove the capping layer, the p$^{++}$-layer as well as parts of the blocking barrier. NiAuGe is deposited on the new surface by electron-beam physical vapour deposition (EBPVD). Low-resistance contacts to the n$^{+}$-layer are formed on thermal annealing. To contact the p$^{++}$-layer, the top-gate, another local etch removes the capping layer. A 100\,nm thick Ti/Au contact pad is deposited on the new surface by EBPVD. This contact is not thermally annealed but nevertheless provides a reasonably low-resistance contact to the top-gate (Fig.~\ref{fig:fig1_detailed}a).

Following the fabrication of the contacts to the back- and top-gates, the contacts themselves are covered with photoresist and the surface of the sample is passivated by chemical treatment. HCl removes a thin oxide layer and a few\,nm of GaAs on the sample surface. After rinsing the sample with deionised water, it is immediately put into an ammonium sulphide ((NH$_4$)$_2$S) bath and subsequently into an atomic layer deposition (ALD) chamber. With ALD, 8\,nm of Al$_2$O$_3$ is deposited at a temperature of 150 $^\circ$C. This process is essential with the present device to reduce surface-related absorption: a high $\mathcal{Q}$-factor is only achieved with a surface passivation layer. We can only speculate on the microscopic explanation at this point. The passivation procedure reduces the surface density-of-states, leading to an unpinning of the Fermi energy at the surface. On the one hand, this reduces the Franz-Keldysh absorption in the capping layer. On the other hand, it reduces the absorption from mid-gap surface states. A clear advantage of the surface passivation is that native oxides of GaAs are removed and prevented from re-forming: this not only reduces the probability for surface absorption but also provides a robust and stable termination to the GaAs sample~\cite{Guha2017}.

A sample holder contains large Au pads. The Ti/Au and NiAuGe films are connected to the Au pads by wire bonding. Silver paint is used to connect the Au pads to macroscopic wires (twisted pairs).

\section{Microcavity design and realisation}
\subsection{Curved top mirror: CO$_2$ laser ablation}
The template for the curved top-mirror is produced by in-house CO$_2$ laser ablation \cite{HungerAIP2012,GreuterAPL2014} on a 0.5\,mm thick fused-silica substrate. The radius of curvature of the indentation is 10.5 $\mu$m as measured by confocal scanning
microscopy \cite{GreuterAPL2014}; the depth relative to the unprocessed surface is 1.2 $\mu$m. After laser ablation, the template is coated with 22 pairs of Ta$_2$O$_5$ (refractive index $n=2.09$) and SiO$_2$ ($n=1.46$) layers (terminating with a layer of SiO$_2$) by ion-beam sputtering~\cite{BeauvilleCQG2004}.

\subsection{Microcavity characterisation}

Each mirror is characterised by measuring the reflection at wavelengths outside the stopband. The reflection oscillates as a function of wavelength. We find that these oscillations are a sensitive function of the exact layer thicknesses of the DBR. The transmission is simulated with a one-dimensional transfer matrix calculation, for instance the Essential Macleod package. A fit is generated, taking the nominal growth parameters as starting point and making the simplest possible assumption to describe systematic differences between the experiment and the calculation. In this way we find that the GaAs (AlAs) layers in the semiconductor DBR start with a physical thickness of 64.6\,nm (80.2\,nm) for $n=3.49$ ($n=2.92$), reducing linearly to 63.9 (79.8\,nm). The change arises simply because the growth rate changes slightly during the long process of growing the DBR. Accordingly, we anticipate that the layers in the active layer have actual thicknesses: n$^{+}$-layer 38.9\,nm; tunnel barrier 29.4\,nm; blocking barrier 183.3\,nm; p$^{++}$-layer 19.0\,nm; p$^{+}$-layer 4.8\,nm; cap 55.8\,nm. The main consequence of the slight change in growth rate during growth is that the stopband centre is shifted from 940\,nm (design wavelength) to 920\,nm. The maximum reflectivity at the stopband centre is not changed significantly by these slight deviations in layer thicknesses.

For technical reasons, the dielectric DBR has a nominal (measured) stopband centre of 1017\,nm (980\,nm), i.e.\ red-detuned from the quantum dot emission. Since the transmission could not be measured during deposition at a wavelength of 940\,nm, a modified quarterwave stack was chosen which is expected to have similar transmission (87\,ppm) at 1064\,nm and 940\,nm. A laser at 1064\,nm was used for \textit{in situ} characterisation. The displacement in stopband centres between top and bottom DBRs is an issue only at wavelengths below 915\,nm where the cavity $\mathcal{Q}$-factor decreases rapidly with decreasing wavelength. Matching of the two stopband centres would give a high $\mathcal{Q}$-factor over a larger spectral range.

A microcavity was constructed using a planar dielectric mirror and the same curved dielectric mirror used for the main quantum dot experiment. Both planar and curved silica templates were coated in the same run. With the smallest possible mirror separation of $3\lambda/2$ (limited by the indentation depth of the curved mirror) we determine $\mathcal{Q}$-factors of $1.7\cdot10^5$ ($1.5\cdot10^6$) at 920\,nm (980\,nm) at room temperature. The fundamental microcavity mode splits into a doublet with orthogonal polarisations. At a wavelength of 920\,nm, this splitting is typically 13\,GHz. These measurements demonstrate the very high quality of the dielectric mirror, in particular the curved dielectric mirror.

The microcavity consisting of the semiconductor mirror and the same curved dielectric mirror has a $\mathcal{Q}$-factor of typically $5\cdot10^5$ at 920\,nm at 4.2\,K (Fig.\ 2, main text), a factor of $\sim3$ larger than the dielectric DBR-dielectric DBR microcavity described above. This increase can be explained by a factor of two larger effective cavity length of the semiconductor-dielectric cavity -- the group delay of the semiconductor mirror is larger than that of a dielectric mirror due to the $3\lambda/2$-thick active layer -- and a factor of 1.5 larger finesse. This increase in finesse suggests that at 920\,nm the reflectance of the semiconductor mirror is higher than that of the dielectric mirror. 

The fundamental mode at wavelength 920\,nm has a polarisation splitting of typically 32\,GHz. This is larger than the polarisation splitting of the dielectric DBR-dielectric DBR microcavity (13\,GHz at 920\,nm). This suggests that the main origin of the polarisation splitting is birefringence in the semiconductor induced by strain (AlAs is not exactly lattice-matched to GaAs).

\subsection{Low-temperature setup and stability} 
Both the top-mirror and the GaAs sample are firmly glued to individual titanium sample holders and mounted inside a titanium ``cage" (Fig.\ \ref{fig:fig1_detailed}a). The holder for the GaAs sample is fixed to a stack of piezo-driven XYZ nano-positioners while the top-mirror holder is fixed to the titanium cage via soft (indium) washers which act as a flexible material for tilt alignment at room temperature. Observing the cavity with a conventional optical microscope and tightening each screw of the mirror holder individually, Newton rings appearing between the two mirrors can be centred in order to guarantee mirror parallelism at room temperature. The entire microcavity setup is then inserted in another titanium cage. This outer cage is connected to an optical cage system inside a vacuum tube. The tube is evacuated, flushed with He exchange gas (25 mbar), pre-cooled in liquid nitrogen and finally transferred into the helium bath cryostat.

In order to minimise the exposure of the microcavity to acoustic noise, the cryostat is decoupled from floor vibrations via both active and passive isolation platforms (Fig.~\ref{fig:fig1_detailed}b). An acoustic enclosure surrounds both the entire cryostat and microscope, providing a shield against airborne acoustic noise (Fig.~\ref{fig:fig1_detailed}b). There is no active feedback mechanism acting on the microcavity's z-piezo. Nevertheless, a root-mean-square cavity length fluctuation~\cite{GreuterAPL2014} of $\sim0.5$ pm was measured in the best case, limiting our $\mathcal{Q}$-factors to $\mathcal{Q}\approx2.0\cdot10^6$. This corresponds to our highest measured $\mathcal{Q}$-factor of $\mathcal{Q}=1.6\cdot10^6$ in the case of a microcavity consisting of the curved top mirror paired with a dielectric bottom mirror of identical coating. This suggests that in the case of a GaAs sample--curved dielectric mirror combination the $\mathcal{Q}$-factor is only slightly reduced by environmental noise.

\section{Quantum dot spectroscopy and second-order correlation measurements}
\subsection{Quantum dot charging}
\begin{figure}[t!]{}
	\centering
	\includegraphics[width=86 mm]{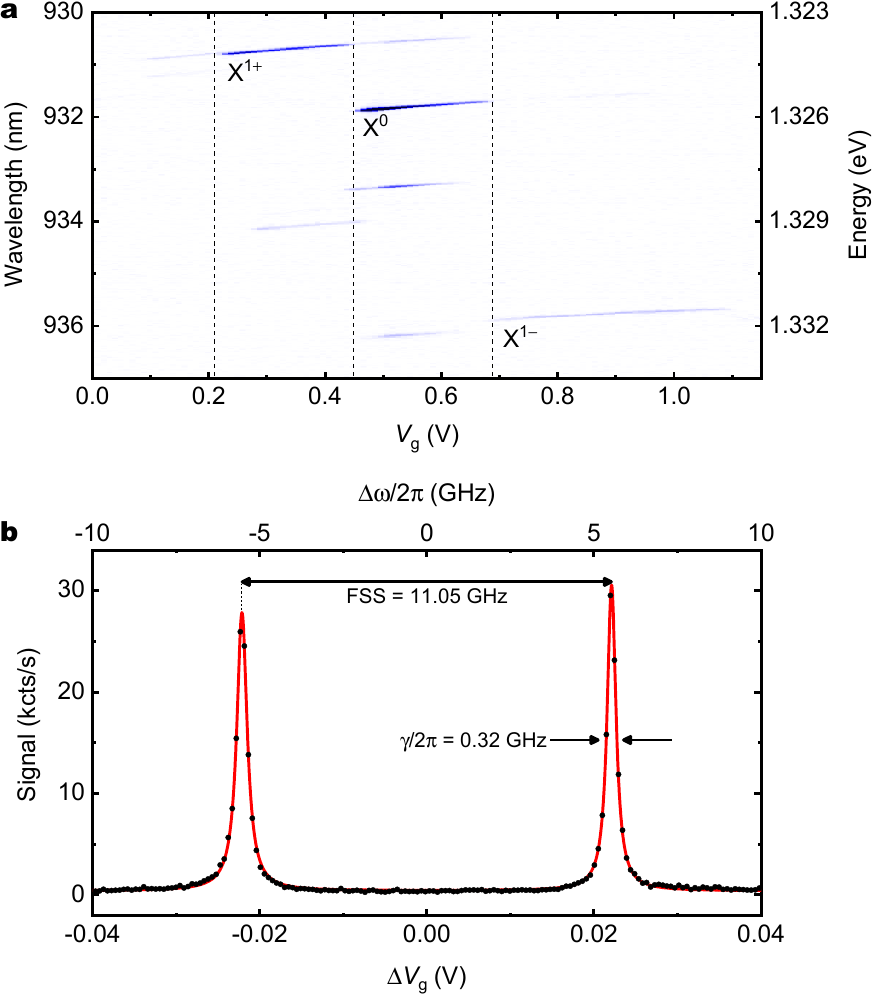}
	\begin{singlespace}
 \caption{
 \textbf{Quantum dot charging and neutral exciton linewidth.}
	\textbf{a} Measured photoluminescence signal of non-resonantly excited QD4 ($\lambda=830$\,nm, $P=200$ nW) as a function of gate voltage. The three main charge states of the quantum dot are the positive trion (X$^+$), neutral exciton (X$^0$) and negative trion (X$^-$). Dark blue: maximum counts, white: minimum counts. \textbf{b} Resonance fluorescence on QD5 (X$^0$, $\lambda=939$\,nm, $B=0.00$ T) excited well below saturation (red solid line: Lorentzian fit). With a measured Stark shift of 240\,GHz/V, a linewidth of 0.32\,GHz is obtained, a value close to the typical transform limit of 0.20\,GHz for these InGAs quantum dots. The splitting arises from the X$^0$ fine structure which for QD5 is 11.05\,GHz.
	}
	\label{fig:supplfigA}
	\end{singlespace}
\end{figure}

To characterise quantum dot charging, photoluminescence (PL) measurements were performed using non-resonant excitation at a wavelength of 830\,nm as a function of the voltage applied between top- and bottom-gates. Fig.~\ref{fig:supplfigA}a shows such a PL charge map taken on the sample without the top mirror. Both positive (X$^+$) and negative (X$^-$) trions as well as the neutral exciton (X$^0$) were identified. The charge states of a quantum dot within the cavity can be recorded in a similar way. In order to detect all the PL before filtering by the cavity, a sine wave voltage is applied to the cavity's z-piezo so that the cavity is continuously scanned through one free spectral range per integration time window of the spectrometer.

\subsection{Resonant excitation: cross-polarised detection of resonance fluorescence}
Each quantum dot's behaviour under resonant excitation can be investigated by suppressing back-reflected laser light in the detection arm, detecting the resonance fluorescence (RF). We achieve this with a dark-field technique \cite{KuhlmannRSI2013}. The optical components are shown in Fig.~\ref{fig:fig1_detailed}b. The excitation laser passes through a linear polariser with polarisation matched to the reflection of the lower polarising beam splitting (PBS). The two PBSs transmit the orthogonal polarisation in the vertical direction, the detection channel. The final polarising element of the excitation channel and the first polarising element of the detection channel is a a quarter-wave plate. It has a dual function. First, by setting the angle of the quarter-wave plate to $\ang{45}$, the microscope can also be operated in bright-field mode. This is very useful for alignment purposes and for optimisation of the out-coupling efficiency. Secondly, in dark-field mode, the quarter-wave plate allows very small retardations to be introduced, correcting for the slight ellipticity in the excitation polarisation state \cite{KuhlmannRSI2013}. The quarter-wave plate allows extremely high bright-field-to-dark-field extinction ratios to be achieved. The microscope can be operated in a set-and-forget mode -- once the polariser and wave-plate are aligned, the laser suppression is maintained over days in the original setup \cite{KuhlmannRSI2013} and even weeks in this case. This very robust operation (despite the fact that control of the wave-plate rotation at the milli-degree level is necessary \cite{KuhlmannRSI2013}) is likely to be a consequence of the effective damping of acoustic and vibrational noise acting on the microscope head in the cavity experiment.

\begin{figure*}[t!]{}
	\centering
	\includegraphics[width=\textwidth]{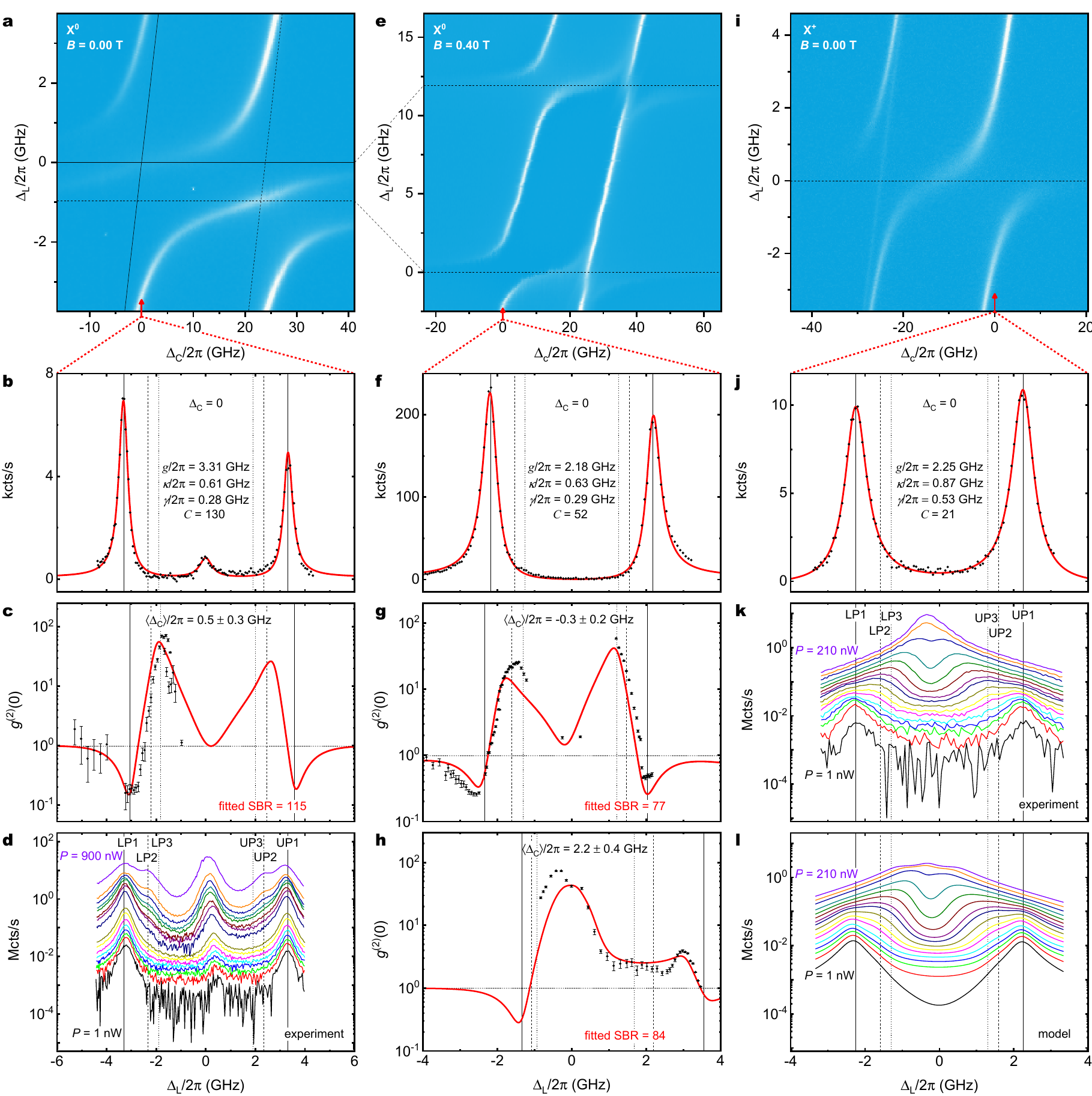}
	\begin{singlespace}
 \caption{
 \textbf{Spectroscopy on cavity-coupled QD1.}
 \textbf{a} X$^0$ at $B=0.00$ T: RF scan revealing two TEM$_{00}$ cavity modes with polarisation splitting of $25$\,GHz (inclined lines) coupled to two fine-structure-split levels of X$^0$ with FSS of 1\,GHz (horizontal lines). \textbf{b} Line cut at resonance to ``left" cavity mode (as indicated by red arrow). The main peaks arise from coupling of the ``high" frequency X$^{0}$ transition to one cavity mode; the peak at $\Delta_\text{L} = 0$ arises from coupling of the ``low" frequency X$^{0}$ transition to the same cavity mode. \textbf{c} $g^{(2)}(0)$ versus laser detuning for a cavity detuning close to zero. \textbf{d} Power dependence at resonance. Excitation of the second rung of the Jaynes-Cummings ladder (LP2, UP2) is evident at high powers as indicated by the dashed vertical lines. \textbf{e} X$^0$ at $B=0.40$ T: RF scan reveals that the same TEM$_{00}$ cavity modes couple to the two X$^{0}$ transitions. The X$^{0}$ transitions are now separated by the Zeeman splitting. \textbf{f} Line cut at resonance to ``left" cavity mode. \textbf{g,h} $g^{(2)}(0)$ versus laser detuning for two different cavity detunings, one close to zero, the other one close to $g$. \textbf{i} X$^{+}$ at $B=0.00$ T: RF scan of the X$^+$ transition. \textbf{j} Line cut at resonance to ``right" cavity mode. \textbf{k,l} Experimental and theoretical power dependence at resonance, respectively. Excitation of higher rungs of the Jaynes-Cummings ladder is evident by the convergence from the two first-rung polaritons towards the bare cavity mode with increasing power. The Hilbert space in the model is truncated to 15 rungs of the Jaynes-Cummings ladder leading to the reduced counts close to the bare cavity mode at high power compared to the experiment. In all figures, the vertical lines depict the resonance frequencies for the first three rungs of the Jaynes-Cummings ladder (LP1/UP1: continuous, LP2/UP2: dashed, LP3/UP3: dotted) at a particular cavity detuning. 
 }
  \label{fig:supplfigB}
 \end{singlespace}
\end{figure*}

\subsection{Second-order correlation measurements and single photon detectors} 
Second-order correlation measurements are performed with a Hanbury Brown-Twiss (HBT) setup. The signal from the detection fibre (Fig.~\ref{fig:fig1_detailed}b) is sent to a 50:50 fibre beam-splitter and then to two superconducting nanowire single photon detectors (SNSPDs, Single Quantum Eos). Each SNSPD has a system detection efficiency of $\approx 85\%$ and a negligible dark count rate (10--40 cts/s). The total timing resolution in the $g^{(2)}$-mode includes the timing resolution of both SNSPDs and the resolution of the time-tagging hardware. In total, it is $\approx 35$ ps (FWHM) which is well below the measured vacuum Rabi periods in this work.

The dead time of the time-tagging hardware is $\simeq95$ ns which sets a limit for the maximally detectable count rate. In order to measure higher count rates than $\sim 5$ Mcts/s per detector, the 1\%-arm of the detection fibre is used instead of the 99\%-arm and the counts are calibrated accordingly. 

For the evaluation of $g^{(2)}(\tau)$ we use a time window of 100 ns. For all presented $g^{(2)}(\tau)$ data, we use a bin size of 4 ps. For all presented $g^{(2)}(0)$ values, we perform a fast Fourier transform (FFT) of $g^{(2)}(\tau)$ (bin size: 16 ps), we then cut all frequency components above 14\,GHz and calculate the inverse FFT. In this way, we make sure that the $g^{(2)}(0)$ values are averaged over a time of 35 ps, a time large with respect to the original binning 16 ps, but small with respect to the period of the vacuum Rabi oscillations.

\subsection{Resonant excitation: neutral exciton} 
An RF scan of QD5 without top mirror is shown in Fig.~\ref{fig:supplfigA}b. The detuning between quantum dot and laser is controlled in this case by fixing the laser frequency and scanning the gate voltage which detunes the quantum dot resonance frequency via the dc Stark shift. Two peaks are observed from the neutral exciton, X$^{0}$. The splitting corresponds to the fine-structure splitting (FSS). Taking several scans for different laser frequencies, a dc Stark shift of 240\,GHz/V is determined on this particular quantum dot. The measured full-width-at-half-maximum of each neutral exciton peak corresponds to 0.32\,GHz, a value close to the transform limit of 0.20\,GHz for these InGaAs quantum dots \cite{KuhlmannNatComm2013}.

\subsection{Resonant excitation: polarisation axes}
The X$^{0}$ polarisation axis (or, in shortened form, ``axis'') varies from quantum dot to quantum dot. The cavity also has an axis. A complication is that the cavity mode splitting (32\,GHz), the X$^{0}$ fine-structure (1--10\,GHz), and the frequency separating the two polaritons in the strong-coupling regime (6--9\,GHz) are all similar. Fig.~\ref{fig:supplfigB}a shows an example: full RF scans of cavity-coupled QD3 are shown, together with their respective line-cuts at zero cavity detuning (Fig.~\ref{fig:supplfigB}b, f, j). The fundamental cavity mode splits into two modes with linear and orthogonal polarisations. At zero magnetic field ($B = 0.00$ T) the neutral exciton X$^{0}$ also splits into two lines with linear and orthogonal polarisations. In the case of QD3 at $B = 0.00$ T, the X$^{0}$ and cavity axes are close-to-parallel such that one X$^{0}$ line couples strongly to one cavity mode, weakly to the other cavity mode, and vice versa for the other X$^{0}$ line (Fig.~\ref{fig:supplfigB}a). The line-cut at one particular cavity frequency shows the polaritons and a weak feature in between (Fig.~\ref{fig:supplfigB}b). The analysis including both cavity modes and two X$^{0}$ transitions makes it clear that in Fig.~\ref{fig:supplfigB}b, the two polaritons arise from strong coupling between one X$^{0}$ transition and one cavity mode. The central feature arises from an out-of-resonance response of the strong coupling between the other X$^{0}$ transition and the other cavity mode. The bare cavity mode is not observed at all in the spectral range of Fig.~\ref{fig:supplfigB}a. 

The quantum dot-cavity couplings can be selected in a few ways in this experiment. 

First, the X$^{0}$ axis varies from quantum dot to quantum dot. It is not difficult to find a quantum dot whose axis matches closely that of the cavity such that one X$^{0}$ line interacts primarily with one cavity mode, the other X$^{0}$ line interacts primarily with the other cavity mode. Fig.~\ref{fig:supplfigB}a depicts an example of this behaviour. 

Secondly, application of a small magnetic field pushes the two X$^{0}$ lines apart in frequency. At a magnetic field of $B=0.40$ T, the X$^{0}$ lines (QD1) are separated by $12$\,GHz such that if one X$^{0}$ line is resonant with the microcavity, the other X$^{0}$ line is far detuned. Fig.~\ref{fig:supplfigB}b,f show an example. At these magnetic fields, the X$^{0}$ lines become circularly polarised such that the X$^{0}$ axis plays no further role. The price to pay is a reduction by a factor of $\sqrt{2}$ in the coupling parameter $g$ with respect to the optimal value at zero magnetic field (Fig.~\ref{fig:supplfigB}f). 

Thirdly, the fine-structure splitting disappears on switching to a charged exciton, either X$^{-}$ or X$^{+}$: there is just one peak at zero magnetic field (Fig.~\ref{fig:supplfigB}i,j), a Zeeman-split doublet at finite magnetic field. 

To exploit all three options, we stress the power of the {\em in situ} cavity detuning. On applying a magnetic field or changing the voltage applied to the device, the quantum dot optical frequency changes by many cavity linewidths but in each case the cavity can be brought into resonance.

\subsection{Resonant excitation: vacuum Rabi frequency versus \boldmath{$\Delta_{\rm C}$}}
\begin{figure}[t!]
\centering
\includegraphics[width=86 mm]{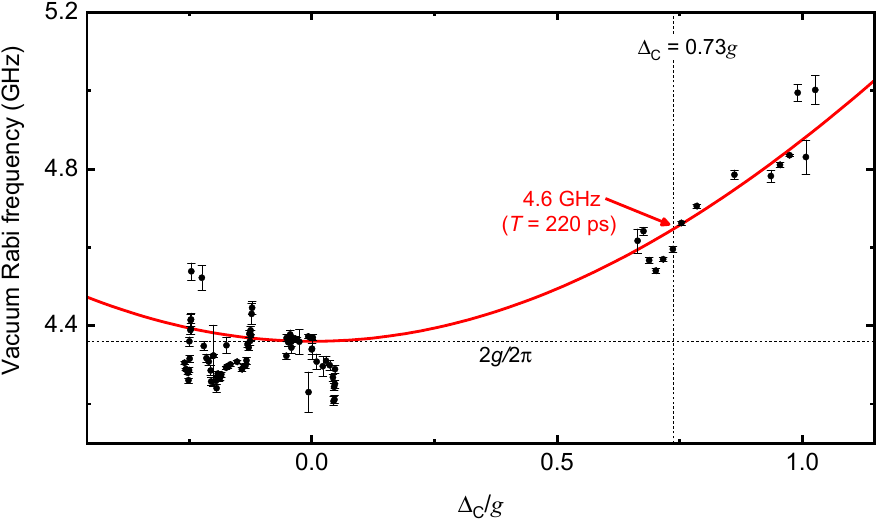}
\begin{singlespace}
 \caption{
 \textbf{Vacuum Rabi frequency versus $\Delta_{\rm C}$.}
 The data points correspond to measured vacuum Rabi frequencies (determined via FFT of $g^{(2)}(\tau)$) for different cavity detunings $\Delta_{\rm C}$. The red solid-line is an analytical calculation of the polariton splitting in the $\ket{1\pm}$ manifold for different values of $\Delta_{\rm C}$ (see Eq.~\eqref{eq:eigenvaluesVSdelC} in Sec.~\ref{sec: two lasers}) using a coupling strength measured via spectroscopy (Fig.~\ref{fig:supplfigB}f). 
 }
  \label{fig:supplfigOsc}
 \end{singlespace}
\end{figure}

Fig.~3 of the main paper shows $g^{(2)}(\tau)$ as a function of delay $\tau$ for a cavity which is detuned by $\Delta_{\rm C}=0.73 g$ with respect to the emitter. Here we show that vacuum Rabi oscillations in $g^{(2)}(\tau)$ are observed for different values of $\Delta_{\rm C}$ and that the frequency of these oscillations changes according to the change in polariton splitting in the $\ket{1\pm}$ manifold for different values of $\Delta_{\rm C}$ (see Fig.~\ref{fig:supplfigOsc} and section V for analytical calculations for the case of $\Delta_{\rm C}=0$). The dashed vertical line in Fig.~\ref{fig:supplfigOsc} depicts the cavity detuning for the data shown in Fig.~3 of the main paper. Consistent with the excellent agreement of the numerical model for $g^{(2)}(\tau)$ with the experiment, an analytical approach to determine the vacuum Rabi period yields $T=220$ ps in exact agreement with the experimental observations.

\begin{figure*}[htbp]{}
	\centering
	\includegraphics[width=\textwidth]{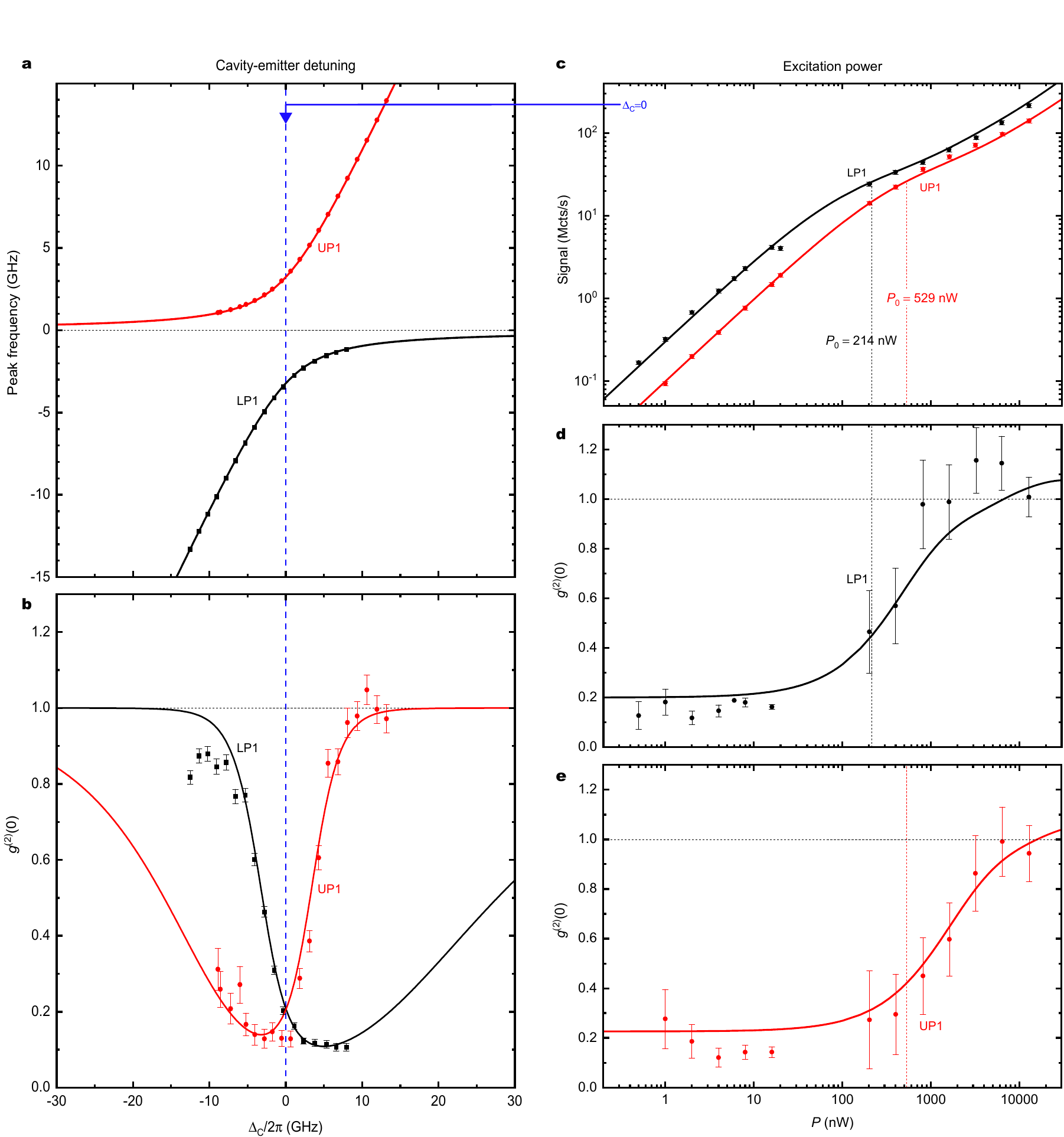}
	\begin{singlespace}
 \caption{
 \textbf{Spectroscopy on cavity-coupled QD2.}
 \textbf{a} Experimental and theoretical dispersion of the lower (LP1) and the upper polariton (UP1). \textbf{b} Corresponding experimental and theoretical $g^{(2)}(0)$ values. \textbf{c} Intensity of scattered light from LP1 and UP1 at zero cavity detuning as a function of resonant excitation power. The absence of saturation is due to population of higher rungs of the Jaynes-Cummings ladder with increasing power. The behaviour at low powers allows the dependence of the Rabi frequency $\Omega$ on laser power $P$ to be determined. This behaviour is parameterised with power $P_{0}$ (see text for definition of $P_{0}$): $P_{0} = 214$ nW for LP1 and $P_{\rm 0} = 529$ nW for UP1 (black and red dashed vertical lines, respectively); \textbf{d,e} corresponding experimental and theoretical $g^{(2)}(0)$ values for LP1 and UP1.
 }
 \label{fig:supplfigD}
 \end{singlespace}
\end{figure*}

\subsection{Resonant excitation: \boldmath{$g^{(2)}(0)$} versus $\Delta_{\rm L}$ and $\Delta_{\rm C}$}
In the experiment, three frequencies can be tuned {\em in situ}: the laser frequency $\omega_{L}$, the emitter frequency $\omega_{C}$ (via the gate voltage) as well as the cavity frequency $\omega_{0}$ (via tuning of the cavity length).

Fig.~4e of the main paper shows $g^{(2)}(0)$ as a function of laser detuning $\Delta_{\rm L}$ for a cavity detuning $\Delta_{\rm C}=0$ on QD2 at $B=0.50$ T. $g^{(2)}(0)$ can be described well with the model and a small laser background. This point is investigated also in other cases. In Fig.~\ref{fig:supplfigB}c,g and h, more $g^{(2)}(0)$ measurements of the neutral excition of QD1 at $B = 0.00$ T and 0.40 T are shown: c and g are recorded with close-to-zero cavity detuning, h with a cavity detuning of $\Delta_{\rm C}\approx g$. 

The {\em in situ} tunability of the microcavity can be exploited by an alternative experiment in which the cavity is detuned and the polaritons are driven resonantly at each cavity detuning. Fig.~\ref{fig:supplfigD}a,b show exactly this, specifically the behaviour of the first-rung polaritons (LP1 in black, UP1 in red) as a function of $\Delta_{\rm C}$. Also in this case, the model reproduces the experimental results well. The reason for the slight discrepancy in $g^{(2)}(0)$ of the lower polariton at large and negative $\Delta_{\rm C}$ is the fact that the laser starts driving the second fine-structure level which is weakly coupled to the same cavity mode. This increases slightly the number of single photons in the detection signal as evidenced by the slight anti-bunching in the experimental data.

\subsection{Resonant excitation: power dependence}
The experiments in Fig.s~1--5 of the main paper and Fig.~\ref{fig:supplfigB}a,b,c,e,f,g,h,i,j and Fig.~\ref{fig:supplfigD}a,b are all recorded with a weak driving laser, i.e.\ with an average photon occupation in the cavity below one. We present here the behaviour as the power of the driving laser increases. 

In Fig.~\ref{fig:supplfigD}c we plot the measured and calculated scattering signal on driving LP1 (black) and UP1 (red) with increasing excitation power. A striking feature is that the system does not saturate (Fig.~\ref{fig:supplfigD}c). This is evidence that the full ladder of Jaynes-Cummings levels exists. To model the power dependence, it is necessary to determine the connection between the Rabi frequency $\Omega$, the input parameter to the model, and the laser power $P$, the control parameter in the experiment. Clearly, $\Omega \propto \sqrt{P}$. At the lowest powers, only the zeroth and first rungs of the Jaynes-Cummings ladder are populated such that the $\ket{0} \leftrightarrow \ket{1-}$ and $\ket{0} \leftrightarrow \ket{1+}$ transitions behave like two-level systems: the scattered signal increases linearly with laser power, as expected (Fig.~\ref{fig:supplfigD}c). We parameterise the link between $\Omega$ and $P$ by adopting the link for a two-level system, namely 
$\Omega=\sqrt{\frac{P}{P_{0}}} \frac{\kappa+\gamma}{2} \frac{1}{\sqrt{2}}$, where $P$ is the laser power (monitored at the 50:50 fibre beam-splitter) and $P_{0}$ is a reference power. Comparing the model to the measured counts and $g^{(2)}(\tau)$ values (plotting the $g^{(2)}(0)$ values only), we fit $P_{0}=214$ nW ($P_{0}=529$ nW) for LP1 (UP1) and an overall detection efficiency of $12\%$.

The difference in powers $P_{0}$ for LP1 and UP1 results in an unequal population of the polaritons at constant input powers, as seen in Fig.~2f,g of the main paper. The difference in $P_{0}$-values probably arises from a polarisation-dependent chromaticity in the throughput of the microscope.

The behaviour as a function of driving power can also be explored by measuring the $\Delta_{\rm L}$-dependence of the scattered intensity for $\Delta_{\rm C}=0$. Fig.~\ref{fig:supplfigB}d,k show power-dependent RF scans when the bare exciton and cavity are resonant. At low power, LP1 and UP1 are clearly resolved. At higher power, bumps appear at the two-photon LP2 and UP2 resonances. In Fig.~\ref{fig:supplfigB}k, there is no resonance close to the bare cavity mode at low power, enabling us to explore the full behaviour even at very large driving powers. At the highest powers, the response is dominated by a feature at $\Delta_{\rm L} \approx 0$ (Fig.~\ref{fig:supplfigB}k). This too is evidence that the full Jaynes-Cummings ladder can be accessed. At the highest powers, the system ``climbs" the Jaynes-Cummings ladder on account of the bosonic enhancement of photons such that the average photon occupation is large and the polariton resonances become closer in frequency to the bare cavity mode.

The power dependence can also be described with the model and very good agreement between our numerical model and the data in Fig.~\ref{fig:supplfigB}k is found. (Due to the presence of the second fine-structure level in Fig.~\ref{fig:supplfigB}d, our numerical model is incomplete in this case.)


\section{Theory: laser driven atom-cavity system}

\subsection{Hamiltonians of the atom-cavity system}

We start by giving the free Hamiltonian of the atom-cavity system. The quantum dot is modelled as a two level system with energy levels $\ket{\t{g}}$ and $\ket{\t{e}}$ separated by an energy $\omega_\t{0}$ (here and in the rest of the section we take $\hbar=1$), i.e. 
\\\be
H_\t{0}= \omega_\t{e} \proj{\t{e}}+ \omega_\t{g} \proj{\t{g}}.
\ee
We set $\omega_\t{e}=\omega_\t{0}$ and $\omega_\t{g}=0$ to simplify the notations. For the cavity, we restrict our consideration to a single mode with associated creation and annihilation operators $a^\dag$ and $a.$ If the frequency of the cavity field is resonant with the frequency separation of the two-level atom, we have
\\\be
H_\text{C} = \omega_\t{0} a^\dag a.
\ee
When the cavity frequency is detuned with respect to the atomic energy, $\omega_0$ has to be replaced by $\omega_\t{C}.$ The interaction between the quantum dot and the cavity mode is described by $g (\ketbra{\t{e}}{\t{g}} +\ketbra{\t{g}}{\t{e}})( a^\dag+a)$, where $g$ is the coupling constant between the quantum dot and the bare cavity mode. In the limit $g \ll \omega_\t{0},$ this coupling Hamiltonian is well approximated by the Jaynes-Cummings Hamiltonian
\\\be
H_\t{int}= g \Big( \ketbra{\t{g}}{\t{e}} a^\dagger + \ketbra{\t{e}}{\t{g}} a \Big).
\ee
The free Hamiltonian of the atom-cavity system is thus given by 
\\\be
H_\t{free}=H_\t{0} + H_\t{C} + H_\t{int}.
\ee

\subsection{Eigenstates and eigenvectors of the atom-cavity system}

\label{sec: hybridizaation}

\begin{figure}[t!]{}
\includegraphics[width=86 mm]{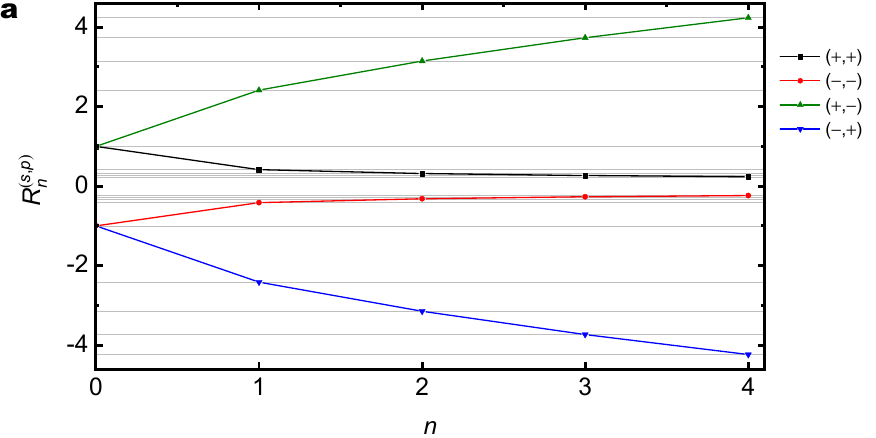}
\begin{singlespace}
\caption{\textbf{Anharmonicity of the Jaynes-Cummings ladder.} Plotted is the deviation of the energy difference between neighbouring ladder rungs (in units of $g$) from the central frequency $\omega_\t{0}$ as a function of excitations $n$ (up to $n=4$) in the atom-cavity system.}
\label{fig:1}
\end{singlespace}
\end{figure}

To simplify the problem, it is convenient to choose a basis where the free Hamiltonian is diagonal. This basis can be easily found by noticing that $H_\t{free}$ is block diagonal with blocks of size two spanned by $\ket{\t{g},n}$ and $\ket{\t{e},n-1}$, and a single block of size one spanned by $\ket{\t{g},0}$ with eigenvalue zero. $\ket{n}$ here denotes the Fock state for the light field with $n$ excitations. Hence, using the basis $\{\ket{\t{g},n},\ket{\t{e},n-1}\}$ for each block we write
\\\be
H_\t{free} = 
\left(\begin{array}{cc}
0 & \\
& \bigoplus_{n=1}^\infty 
\left[\begin{array}{cc}
 n \omega_\t{0} & \sqrt{n} g \\
 \sqrt{n} g & n \omega_\t{0}
\end{array}\right]
\end{array}\right),
\ee
which can be easily diagonalised. The eigenstates of $H_\t{free}$ are given by
\\\be
\ket{n\pm}= \frac{\ket{\t{g},n}\pm \ket{\t{e},n-1}}{\sqrt{2}},
\ee
with energies
\\\be
E_n^{\pm} = n \omega_\t{0} \pm \sqrt{n} g, 
\ee
for $n \geq 1$ and $\ket{0} = \ket{\t{g},0}$ with $E_0=0$. As a result,
\\\be
H_\t{free}= \sum_{n,\pm} E_n^{\pm} \proj{n\pm}.
\ee
To illustrate the anharmonicity of its spectrum, we plot in Fig.~\ref{fig:1} the deviation of the energy difference between the neighbouring levels from the central frequency $\omega_\t{0}$,
\\\be
R_n^{(s,p)} = \frac{E^s_{n+1} - E^p_n - \omega_\t{0}}{g},
\ee
expressed in units of $g$.

One notes that in the case where the cavity mode frequency $\omega_\t{C}$ is not exactly equal to the atomic frequency $\omega_\t{0}$ the free Hamiltonian reads
\begin{equation}\label{eq: H original}
H_\t{free} = 
\left(\begin{array}{cc}
0 & \\
& \bigoplus_{n=1}^\infty 
\left[\begin{array}{cc}
 n \omega_\t{C} & \sqrt{n} g \\
 \sqrt{n} g & n \omega_\t{C} +(\omega_\t{0}-\omega_\t{C})
\end{array}\right]
\end{array}\right),
\end{equation}
which affects both the spectrum and the eigenbasis, as we shall see later.

\subsection{Master equation of the atom-cavity system driven by laser light}
We consider the case where the atom-cavity system is driven by a laser through 
\begin{equation}\label{eq: hamiltoniann}
H_\t{L}(t) = \Omega(t) a + \Omega^\star(t) a^\dag
\end{equation}
where $\Omega(t)$ is proportional to the complex light field amplitude incident on one of the cavity mirrors at time $t$. The cavity photons can leak out of the cavity via a beam-splitter type interaction. This leads to a decay channel entering in the master equation via a Lindblad operator $ L_\kappa= \sqrt{\kappa} a$. Similarly, the spontaneous decay of the level $\ket{\t{e}}$ to $\ket{\t{g}}$ appears in the master equation via $L_\gamma = \sqrt{\gamma} \ketbra{\t{g}}{\t{e}}$. The evolution of this driven system is thus given by the following master equation 

\begin{equation}\label{eq:ME}
\dot \rho = -\ii [H_\t{tot},\rho] +\!\!\!\! \sum_{L =L_\kappa, L_\gamma}\!\!\!\!\left(L \rho L^\dag - \frac{1}{2} L^\dag L \rho -\frac{1}{2} \rho L^\dag L \right)
\end{equation}
with $H_\t{tot} = H_\t{free}+H_\t{L}(t).$

\subsection{Total Hamiltonian and Lindblad operators in the eigenbasis of the free Hamiltonian}

We can express the bosonic operator $a$ in the basis $\{\ket{n\pm}\}$ as
\\\begin{align}\label{eq: a in hb}
a = &\ket{0}\frac{\bra{1+}+\bra{1-}}{\sqrt{2}} +\nonumber\\
&\sum_{n\geq1} \nonumber
(\ket{n+}\, \ket{n-} )
\left(\begin{array}{cc}
T_n^+ & T_n^- \\
T_n^- & T_n^+
\end{array}\right) 
\left(\begin{array}{c}
\bra{(n+1)+}\\
\bra{(n+1)-}
\end{array}\right) \\
=&\left(\begin{array}{c|cc|c|cc|c}
0& \frac{1}{\sqrt{2}} & \frac{1}{\sqrt{2}} & & & &\\
\hline
& & &\dots & & &\\
\hline
& & & &T_n^+&T_n^-&\\
& & & &T_n^-&T_n^+&\\
\hline
& & & & & &\dots
\end{array}\right)
\end{align}
with $T_n^\pm = \frac{\sqrt{n+1}\pm \sqrt{n}}{2}$.
Similarly, we have
\\\begin{align*}
 \ketbra{\t{g}}{\t{e}} =
 &\ket{0}\frac{\bra{1+}-\bra{1-}}{\sqrt{2}} +\nonumber\\
&\sum_{n\geq1}
(\ket{n+}\, \ket{n-} )
\left(\begin{array}{rr}
\frac{1}{2} & -\frac{1}{2} \\
-\frac{1}{2} & \frac{1}{2}
\end{array}\right) 
\left(\begin{array}{c}
\bra{(n+1)+}\\
\bra{(n+1)-}
\end{array}\right).
\end{align*}
This means that both the total Hamiltonian and the Lindblad operators can be conveniently expressed in the basis $\{\ket{n\pm}\}.$

\subsection{Vectorisation of the master equation} 

The master equation is linear and the most direct way to solve it is to write it down in a vector form. One can express an arbitrary density matrix 
\\\be
\rho = \sum_{ij} \rho_{ij} \ketbra{i}{j}
\ee
as
\\\be
v = \t{vec}(\rho) = \sum \rho_{ij} \ket{i}\ket{j}.
\ee
Then it is easy to see that 
\\\be
\t{vec}(-\ii [H, \rho]) = u[H]v
\ee
where 
\\\be
u[H] = -\ii \left(H\otimes \Id - \Id \otimes H^T\right)
\ee
and
\\\be
\t{vec}(L \rho L^\dag -\frac{1}{2}\{L^\dag L , \rho\}) = \ell[L] v
\ee
with
\\\be
\ell[L] = (L\otimes L^* -\frac{L^\dag L}{2} \otimes \Id - \Id \otimes \frac{(L^\dag L)^T}{2}).
\ee
In this representation the master equation~\eqref{eq:ME} takes a simple form
\\\be
\dot v = \underbrace{\left(u[H_\t{tot}]+ \ell[L_\kappa]+ \ell[L_\gamma]\right)}_D v.
\ee
For convenience, we can also define the inverse transformation $\rho = \t{vec}^{-1}(v)$ which simply arranges the components of $v$ in a matrix form. The solution of the master equation~\eqref{eq:ME} can thus formally be written as $\textrm{e}^{\mathcal{L}t}[\rho] = \t{vec}^{-1} [e^{Dt}. \t{vec}(\rho)]$ when $D$ is time independent, $\mathcal{L}$ being the Lindblad superoperator.

\subsection{Numerical solutions for the dynamics of the atom-cavity system driven by a single monochromatic laser}
Analytical solutions of the master equation can be found by focusing on the relevant atom-cavity energy states and discarding the remaining states (Section \ref{analytics}). The truncation of the Hilbert space in the analytical models is less severe in a fully numerical model. We make use of the Quantum Toolbox in Python (QuTiP)\cite{qutip2} and truncate the Hilbert space to $n=15$ in order to model the experimental results. (This numerical model is referred to as ``model'' in all figure legends of this work.) We consider the case where the quantum dot-cavity system is driven by a single monochromatic laser with frequency $\omega_L$. Eq.~\eqref{eq: hamiltoniann} becomes 
\begin{equation}\label{eq: laserhamiltonian}
H_\t{L}(t) = \Omega e^{\ii \omega_\t{L} t } a + \Omega e^{-\ii \omega_\t{L} t } a^\dag.
\end{equation} 
As $\Omega$ is time independent, we can eliminate the explicit time dependence of the total Hamiltonian by considering the frame rotating at $\omega_L$:
\\\begin{align*}
H_{\t{rf}} =&\Omega\left( a + a^\dagger \right) + (\Delta_\t{C}-\Delta_{\t L}) a^\dagger a - \Delta_\t{L} \ketbra{\t{e}}{\t{e}} \nonumber\\
&+ g\Big( \ketbra{\t{g}}{\t{e}} a^\dagger + \ketbra{\t{e}}{\t{g}} a \Big).
\end{align*}
The laser detuning relative to the emitter is denoted by $\Delta_\t{L}=\omega_\t{L}-\omega_\t{0}$, while the cavity detuning is denoted by $\Delta_\t{C}=\omega_\t{C}-\omega_\t{0}$. To simulate the experimental results the procedure outlined in references\cite{HamsenPRL2017,FischerPRA2017} is followed. First, we define the two collapse operators which determine the decay to the environment, i.e.\ out of the Jaynes-Cummings system. These are the two Lindblad operators $L_\kappa$ and $L_\gamma$ for the decay process out of the cavity mode and the decay of the quantum dot into leaky modes, respectively. Next, exploiting the quantum regression theorem, one can solve for the normalized second order correlation function, 
\\\be
g^{(2)}(\tau)=\frac{\langle a^\dagger a^\dagger(\tau)a(\tau)a\rangle}{\langle a^\dagger a\rangle^2}=\frac{\textrm{Tr}(a^\dagger a \textrm{e}^{\mathcal{L}\tau}[a\rho_{*} a^\dagger])}{\left(\textrm{Tr}( a^\dagger a\rho_{*})\right)^2},
\ee
where $\mathcal{L}$ is the Lindblad superoperator and $\rho_{*}$ is the steady-state solution of the master equation.\\

In the experiments, even a small amount of mixing of the signal from the quantum dot-cavity system with a laser background can play an important role. This can be included in the model via a simple beam-splitter. The creation and annihilation operators for the input and output modes are denoted $a/b$ and $c/d$, respectively. The transmission and reflection coefficients are given by $t=t'=\sqrt{\eta}$ and $r=r'=i\sqrt{1-\eta}$.
\\\be
\begin{pmatrix}
\hat{c}^{\dagger}\\
\hat{d}^{\dagger}
\end{pmatrix}=\hat{U}_{\rm bs}\begin{pmatrix}
\hat{a}^{\dagger}\\
\hat{b}^{\dagger}
\end{pmatrix},
\ee
\\\be
\hat{U}_{\rm bs}=\begin{pmatrix}
t' & r\\
r' & t
\end{pmatrix}=\begin{pmatrix}
\sqrt{\eta} & i\sqrt{1-\eta}\\
i\sqrt{1-\eta} & \sqrt{\eta}
\end{pmatrix}.
\ee
The creation and annihilation operators of output mode $c$ are given by
\\\be
\hat{c}^{\dagger}=\sqrt{\eta}~\hat{a}^{\dagger}+i\sqrt{1-\eta}~\hat{b}^{\dagger}
\ee
\\\be
\hat{c}=\sqrt{\eta}~\hat{a}-i\sqrt{1-\eta}~\hat{b}
\ee
and similarly for the output mode $d$. Input mode $b$, which is introduced to model the laser background, is in a coherent state with an average photon number, $|\alpha|^2$. 

Experimentally, it is straightforward to determine what percentage of the observed count rate is due to the laser background. By choosing $|\alpha|^2$ to be the photon number expectation value when the system is driven on resonance, the mixing parameter $1-\eta$ determines what percentage of counts comes from the laser background and its inverse value corresponds to the signal-to-background ratio (SBR). It should be noted that this is a good way of including the background in the low excitation regime where the relationship between incident power and count rate is constant such that $\eta$ remains constant.

In order to model background counts (which are proportional to the incoming laser power) for the entire power range (e.g. for calculations in Fig.~\ref{fig:supplfigD}c,d,e), we use a fixed $\eta=0.999$ and a second beam-splitter before input mode $b$ with transmission $t_2=t_2'=\sqrt{\eta_2}$. We choose $\alpha_2=\sqrt{1000 P \eta_2}$ where $P$ is the monitored laser power in the excitation arm of the microscope.

\section{Theory: Analytical solution for the dynamics of the atom-cavity system driven by a single monochromatic laser}
\label{analytics}
The numerical model includes a large number of rungs of the Jaynes-Cummings ladder and gives extremely reliable results particularly at low driving powers when only the first few rungs of the Jaynes-Cummings ladder are populated. However, it does not provide any deep insights. To complement the numerical model, we therefore present analytical calculations. To ensure that the problem is tractable, the Hilbert space has to be highly truncated. As such, the results are not as exact as those from the numerical model. However, the analytical model provides us with the insights which otherwise are missing.
\subsection{Time-independent Hamiltonian and master equation}
As before, we here consider the case where the atom-cavity system is driven by a single monochromatic laser with frequency $\omega_L$, see Eq.~\eqref{eq: laserhamiltonian}.

We retain the basis $|n_\pm\rangle$ and since $\Omega$ is time independent, we can eliminate the explicit time dependence of the total Hamiltonian by considering the rotating frame 
\\\begin{align*}
\ket{n\pm} \to e^{\ii n\omega_L t} \ket{n\pm}.
\end{align*}
In this frame the total Hamiltonian becomes 
\\\begin{align*}
H_\t{rf} = \Omega \,(a + a^\dag) + H_\t{free} -\sum_{n,\pm} n\, \omega_\t{L} \proj{n\pm},
\end{align*}
where the explicit time dependence of the basis vectors leads to the energy shifts given by the last term. Defining the laser detuning as $\Delta_\t{L}=\omega_\t{L} -\omega_\t{0}$ and considering a resonant cavity field, the total Hamiltonian can be written as
\begin{empheq}[box=\widefbox]{align}\label{eq: H rf}
H_\t{rf} = \Omega \,(a + a^\dag) + \sum_{n,\pm} \underbrace{\left(\pm \sqrt{n} g - n \Delta_\t{L} \right)}_{\Delta_n^\pm} \proj{n\pm}
\end{empheq}
where the bosonic operators $a$ and $a^\dag$ can themselves be written as defined in Eq.~\eqref{eq: a in hb}. We deduce that the dynamics of the atom-cavity system driven by a single monochromatic laser is given by
\begin{equation}\label{eq: ME rf}
\dot \rho = -\ii [H_\t{rf},\rho] +\!\!\!\! \sum_{L=L_\kappa, L_\gamma}\!\!\!\!\left(L \rho L^\dag - \frac{1}{2} L^\dag L \rho -\frac{1}{2} \rho L^\dag L \right).
\end{equation}
Note that the operators $L_\kappa$ and $L_\gamma$ have an explicit time dependence $e^{-\ii \omega_L t}$ in the rotating frame, but because they enter as noise this phase does not appear in the master equation (but it takes care of the phase of emitted photons). 
As discussed before, Eq.~\eqref{eq: ME rf} can be written as 
\\\be
\dot v = \left(u[H_\t{rf}]+ \ell[L_\kappa]+ \ell[L_\gamma]\right) v.
\ee
The steady-state $v_*=\t{vec}(\rho_*)$ is given by 
\begin{equation}\label{eq: steady}
\left(u[H_\t{rf}]+ \ell[L_\kappa]+ \ell[L_\gamma]\right) v_* = 0.
\end{equation}

\subsection{Single photon emission rate}
The Hilbert space of the atom-cavity system is in principle infinite. However, for a fixed laser power one expects that for a large enough value of $n,$ the levels $\ket{n\pm}$ have a negligible population in the steady state and more generally do not influence the dynamics of the system. It is not clear a priori at which $n$ one can truncate the Hilbert space. However, once one finds the steady state of the truncated system, it can be verified that the population in the levels $\ket{n\pm}$ is close to zero, a necessary condition for the truncation at $n$ to be meaningful.

For a given $n$, the steady-state $ \rho_* =\t{vec}^{-1}(v_*)$ can be found from Eq.~\eqref{eq: steady}. Such a state can be used to compute the emission rate of photons from
\\\be
p_s = \tr{L_\kappa \rho_* L_\kappa^\dag }.
\ee
The state right after the photon emission is given by
\\\be
\rho|_s = \frac{1}{p_s} L_\kappa \rho_* L_\kappa^\dag,
\ee
or equivalently in the vector representation $v|_s= \frac{1}{p_s} (L_k \otimes L_k^*) v_*.$

\subsection{Conditional emission rate}
For the autocorrelation measurement, we are interested in the probability of detecting a second photon with some time delay $\tau$ after a first emission. To obtain this probability, we can first compute the state of the system at time $\tau$ after the emission of a first photon. In the vector representation, it is simply given by
\\\be
v|_s{(\tau)} = e^{\left(u[H_\t{rf}]+ \ell[L_\kappa]+ \ell[L_\gamma]\right) \tau} v|_s.
\ee
and the emission rate of a second photon at this time is simply given by
\\\be
p_{|s}(\tau) = \tr{L_\kappa \rho|_s^{(\tau)}L_\kappa^\dag}
\ee
with $\rho|_s^{(\tau)} = \t{vec}^{-1}(v|_s{(\tau)}).$ By computing the single photon emission rate and the conditional emission rate, we have all the necessary ingredients to model the result of autocorrelation measurements. The next section gives analytical expressions for a laser frequency on resonance with $\ket{2\pm}$ through a two-photon process.

\subsection{Resonant two-photon excitation}
In this section, we consider the situation where the laser frequency is such that one of the two photon detunings $\Delta_2^\pm$ is small. We further consider the case where the Rabi frequency $\Omega$ is small enough so that we can describe the system within the truncated Hilbert space $\cH=\t{span}\{\ket{0},\ket{1+},\ket{1-},\ket{2\pm} \}$. Inside $\cH,$ the evolution is generated by the Hamiltonian
\\\be H=
\left(\begin{array}{c|cc|c}
0 &\frac{\Omega}{\sqrt{2}} &\frac{ \Omega}{\sqrt{2}} &\\
\hline
\frac{ \Omega}{\sqrt{2}}& \Delta_{1}^+ & & \Omega T_1^{\pm}\\
\frac{\Omega}{\sqrt{2}} & & \Delta_{1}^- & \Omega T_1^{\mp}\\
\hline & \Omega T_1^{\pm}& \Omega T_1^{\mp}& \Delta_2^\pm
\end{array}\right).
\ee 
Since we focus on the regime $\Delta_2^\pm, \Omega \ll \Delta_{1}^\pm$, we can treat the part of the Hamiltonian depending on $\Omega$ and $\Delta_2^\pm$ as a perturbation. To first order in $\Omega$ and $\Delta_2^\pm,$ the dynamics of $|0\rangle$ and $|2\pm\rangle$ is decoupled from the one of $|1+\rangle$ and $|1-\rangle$ and the relevant Hamiltonian for the two-photon transition is given by
\\\be\label{eq:Heff}
H|_{0,2\pm} =
\Omega^2 \left(
\begin{array}{c|c}
\frac{1}{2\Delta_{1}^+} + \frac{1}{2\Delta_{1}^-}& \frac{T_1^\pm}{\sqrt{2}\Delta_{1}^+}+\frac{T_1^\mp}{\sqrt{2}\Delta_{1}^-} \\
\hline
\frac{T_1^\pm}{\sqrt{2}\Delta_{1}^+}+\frac{T_1^\mp}{\sqrt{2}\Delta_{1}^-} & \frac{(T_1^\pm)^2}{\Delta_{1}^+} + \frac{(T_1^\mp)^2}{\Delta_{1}^-} + \frac{\Delta_2^\pm}{\Omega^2}
\end{array}\right).
\ee
The effective detuning 
\\\begin{align*}
\Delta_\t{eff} &= \Omega^2\left(\frac{(T_1^\pm)^2}{\Delta_{1}^+} + \frac{(T_1^\mp)^2}{\Delta_{1}^-}- \frac{1}{2\Delta_{1}^+} - \frac{1}{2\Delta_{1}^-}\right) + \Delta_2^\pm
\end{align*}
includes a light shift given by $\approx \pm \frac{5 \Omega^2}{\sqrt{2} g}.$ It can be set to zero by the choice of the laser frequency $\omega_L^\pm = \frac{1}{2}(2\omega_\t{0}\pm \sqrt{2}g \pm \frac{5 \Omega^2}{\sqrt{2} g}).$ In this case, we obtain an effective Hamiltonian taking a simple form 
\\\be\label{eq:Heff2}
H_\t{eff} \approx
\left(
\begin{array}{c c}
0 & \Omega_\t{eff} \\
\Omega_\t{eff} & 0
\end{array}\right).
\ee
where $\Omega_\t{eff} \approx 2\sqrt{2}\Omega^2/g.$
\subsubsection{Emission rate of the first photon}
\label{sec: signal}
We have obtained an effective Hamiltonian, but to solve the dynamics we also have to consider the noise terms. Both noise processes $L_\kappa$ and $L_\gamma$ are responsible for the decay of the state $\ket{2\pm}$ to the $\{\ket{1+},\ket{1-}\}$ manifold. The rate of such a decay is given by 
\\\be
\kappa_\t{eff} = \tr{ (L_\kappa^\dag L_\kappa + L_\gamma^\dag L_\gamma) \proj{2\pm} } = \frac{3}{2}\kappa + \frac{1}{2}\gamma.
\ee
Once the atom decays into the $\ket{1\pm}$ manifold, it becomes transparent to the laser until it falls to $\ket{\t{g},0}$ after some time. Hence, to find the steady state, we can replace the noise terms by an effective noise
\\\be
L_\t{eff} = 
\left(
\begin{array}{c|c}
\quad & \sqrt{\kappa_\t{eff}} \\
\hline
 & 
\end{array}\right).
\ee
We now have all the elements to write the master equation associated with the effective dynamics
\\\begin{align*}\label{eq: ME s}
\dot \rho = - \ii 
[H_\t{eff},\rho] + L_\t{eff}\rho L_\t{eff}^\dag - \frac{1}{2} \{\rho, L_\t{eff}^\dag L_\t{eff}\} .
\end{align*}
This master equation can be easily solved and the steady-state solution is given by
\\\be\label{eq: rho statr 2}
\rho_* =\frac{1}{2}(\sigma_0 - \frac{4 \Omega_\text{eff} \kappa_\t{eff}}{8 (\Omega_\text{eff} )^2+(\kappa_\t{eff})^2 }\sigma_y + \frac{(\kappa_\t{eff})^2 }{8 (\Omega_\text{eff} )^2+(\kappa_\t{eff})^2} \sigma_z ),
\ee
where $\sigma_0=\ketbra{\t{g}}{\t{g}}+\ketbra{2\pm}{2\pm}$, $\sigma_y=-i\ketbra{\t{g}}{2\pm}+i\ketbra{2\pm}{\t{g}}$ and $\sigma_z=\ketbra{\t{g}}{\t{g}}-\ketbra{2\pm}{2\pm}$. Hence, the emission rate $p_\t{s} =\tr{L_\kappa \rho_* L_\kappa^\dag}$ for the first photon reads 
\begin{empheq}[box=\widefbox]{align}\label{eq: ps}
p_\t{s} = \frac{ 3\, \kappa }{4} \left(1- \frac{1}{1+8 (\Omega_\text{eff}/\kappa_\t{eff})^2}\right).
\end{empheq}
For short enough detection windows $\tau_\t{det}$, the probability that a first photon is emitted is given by $p_\t{s} \tau_\t{det}.$ \\

\subsubsection{Emission rate of the second photon}
\label{sec: idler}
The detection of a photon at time $t_s=0$ leaves the system in the state 
\\\be\label{eq:state 0 gsi}
\ket{\psi^{\pm}|_s} = \frac{a \ket{2\pm}}{||a \ket{2\pm}||} = \frac{\sqrt{2}}{\sqrt{3}}\ket{\t{g},1} \pm \frac{1}{\sqrt{3}}\ket{\t{e},0}.
\ee
We now compute the emission rate for the second photon after a time delay $t$. To do so we have to solve a specific master equation which accounts for three processes: 
\begin{itemize}
\item The unitary evolution is governed by a Hamiltonian which can be written in the $\{\ket{\t{g},1}, \ket{\t{e},0}\}$ subspace as
\\\be
H=\left(\begin{array}{cc}
\omega_\t{0} & g\\
g & \omega_\t{0}
\end{array}\right).
\ee
\item Spontaneous emission by the atom (outside of the cavity mode) is given by Lindblad operator $L_\gamma = \sqrt{\gamma}\,\sigma_-$. In our subspace it takes the form $L_\gamma = \sqrt{\gamma}\,\ketbra{\t{g},0}{\t{e},0}$, with 
\\\be
L_\gamma^\dag L_\gamma= \gamma \proj{\t{e},0}.
\ee
It brings the state outside of the subspace $\{\ket{\t{g},1}, \ket{\t{e},0}\}$, so for simplicity we will simply ignore the terms $L_\gamma \rho L_\gamma^\dag$, and only keep the terms which lead to the decay of probability
\\\be
\bar{ \mathcal{L}}_\gamma = - \frac{1}{2} \{L_\gamma^\dag L_\gamma,\rho \}.
\ee
\item
The emission of the photon outside the cavity is governed by $\sqrt{\kappa} a$, which in our subspace reads $L_\kappa = \sqrt{\kappa} \ketbra{\t{g},0}{\t{g},1}$ with
\\\be
L_\kappa^\dag L_\kappa= \kappa \proj{\t{g},1}.
\ee
As above we only keep the term 
\\\be
\bar{ \mathcal{L}}_\gamma = - \frac{1}{2} \{L_\kappa^\dag L_\kappa,\rho \}.
\ee
\end{itemize}
Hence the master equation is given by 
\\\be\label{eq:ME gsi}
\dot \rho = - \ii [H,\rho] - \frac{1}{2}\{ L_\kappa^\dag L_\kappa + L_\gamma^\dag L_\gamma,\rho\}.
\ee
We expend $\rho$ in the Pauli basis 
\\\be
\rho = \sum_{\alpha =0,x,y,z} n_\alpha \sigma_\alpha.
\ee 
One easily obtains the components ${\bf v}=(n_0, n_x, n_x, n_z)$ as a function of time (and hence the time-dependent state $\rho_t^\pm$) from the initial state with corresponding vector ${\bf v}|_s^{\pm}=(\frac{1}{2},\pm\frac{\sqrt{2}}{3},0,\frac{1}{6})$ using ${\bf v}_t^{\pm} = e^{t M} {\bf v}|_s^\pm$ where 
\\\be
\nonumber
M = 
\left(
\begin{array}{cccc}
 \frac{1}{2} (\gamma +\kappa ) & 0 & 0 & \frac{\gamma -\kappa }{2} \\
 0 & \frac{-1}{2} (\gamma +\kappa ) & 0 & 0 \\
 0 & 0 & \frac{-1}{2} (\gamma +\kappa ) & -2 g \\
 \frac{\gamma -\kappa }{2} & 0 & 2 g & \frac{-1}{2} (\gamma +\kappa ) \\
\end{array}
\right).
\ee
Importantly, the emission rate of the second photon at 
time $t$ is given by $p_i^\pm(t)=\tr{ L_\kappa^\dag L_\kappa \rho_t^\pm}$, with 
\\\be
L_\kappa^\dag L_\kappa =
\left(\begin{array}{cc}
\kappa & 0 \\
0 & 0
\end{array} 
\right)
= \kappa \frac{\sigma_0 + \sigma_z }{2},
\ee
such that $\tr{ L_\kappa^\dag L_\kappa \rho_t} = 2 \, (\frac{\kappa}{2}\, 0\, 0 \, \frac{\kappa}{2})\cdot {\bf v}_t$ (recall that $\tr{\sigma_\alpha^2}=2$). Consequently $
p_i^{\pm} (t) = \kappa \big(n_0(t) +n_z(t) \big) = \kappa (1\, 0\, 0\, 1) \,e^{t M} \,{\bf v}|_s^{\pm}.$
Straightforward algebra shows that $p_i^{+} (t) = p_i^{-} (t) = p_i(t)$ where
\\\begin{align}\label{eq: pi 1}
p_i(t) = \frac{\kappa\, e^{-t \frac{\kappa+\gamma}{2}}} {24 B^2}&(12 g^2+4(4B^2-3g^2)\cos (2 B t) \nonumber\\
&+4 B (\gamma -\kappa ) \sin (2 B t))
\end{align} 
with $B =\sqrt{ g^2 - (\frac{\gamma-\kappa}{4})^2}.$ This can be conveniently rewritten as
\begin{empheq}[box=\widefbox]{align}\label{eq: pi}
p_i(t)= &\frac{\kappa g \,e^{-t \frac{\kappa+\gamma}{2}}}{12 B^2}( 6 g\\ \nonumber
&+ \sqrt{4 g^2 +2 (\gamma-\kappa)^2 } \cos(2 B t+\varphi) ),
\end{empheq}
where $\varphi$ is some phase-shift of the oscillations at $t=0$. In particular $p_i(0) = \frac{2 \kappa}{3}$.

\subsubsection{Time dependence of the autocorrelation measurement}

For short detection intervals the probability of a coincidence with a delay $t$ is proportional to $p_2 (t) = p_s p_i(t)$ such that
\\\be
g^{(2)}(\tau) = \frac{p_2(\tau)}{p_s^2} = \frac{p_i(\tau)}{ p_s}.
\ee
Combining Eqs.~\eqref{eq: ps} and \eqref{eq: pi 1} one easily finds for the autocorrelation for zero delay
\\\be
g^{(2)}(0)= \frac{8}{9}+ \frac{g^2(\gamma+3 \kappa)^2}{288 \Omega^4}.
\ee
From Eq.~\eqref{eq: pi}, we see that $g^{(2)}(\tau)$ has an envelope decaying as $e^{-\tau\frac{\kappa+\gamma}{2}}.$ It also oscillates with a frequency $\sqrt{4 g^2 - (\gamma-\kappa)^2}$ due to rotations of the state in the $\ket{1\pm}$ manifold. In the experiment, $4g^2\gg(\gamma-\kappa)^2$, which explains the origin of the observed oscillation with a frequency $2g$.

\subsection{Measuring the autocorrelation function}
Before presenting theoretical results for the case with two lasers, we would like to compare the way the autocorrelation is computed theoretically and measured in an experiment. The value of $g^{(2)}(0)$ is computed from
\\\be
g^{(2)}(0)= \frac{\tr{ \rho\, a^{\dag 2} a^2 }}{\left(\tr{\rho \, a^\dag a}\right)^2}.
\ee

In our experiment, however, $g^{(2)}(0)$ is measured with two non-photon number resolving (NPNR) detectors after a $50/50$ beam-splitter. Let us label $a$ the input mode and $d_1$ and $d_2$ the modes after the beam-splitter, that is, the detection modes.

For the detector measuring the mode $d_1,$ the positive operator valued probability measure (POVM) element corresponding to a non-detection event is given by 
\begin{equation}\label{eq:p n d}
p_\t{n.d.} = (1-T)^{d_1^\dag d_1}.
\end{equation}
Here, $T$ is the probability that a cavity photon triggers a detection event, $T= \kappa \tau_\t{det} \eta_\t{det} $ with the detection window $\tau_\t{det}$ and the overall detection efficiency $\eta_\t{det}$. This expression \eqref{eq:p n d} can be simply understood as the probability that none of the $d_1^\dag d_1$ cavity photons are detected. So the POVM element associated to the event ``click" is $1 - p_{n.d.}$. If such a detector is preceded by a $50/50$ beam-splitter, the latter simply has the effect of reducing the efficiency by half when acting on the mode before the beam-splitter. Hence, the probability to obtain a detection can be computed directly from the state of mode $a$ from
\\\be
p_{s1} = p_{s2} = 1-\left(1-\frac{T}{2}\right)^{a^\dag a}.
\ee
Similarly, the probability of a coincidence (both detectors to click) is given by 
\\\be
p_\t{C} = \left(1- (1-T)^{d_1^\dag d_1}\right)\left(1- (1-T)^{d_2^\dag d_2}\right),
\ee
where we emphasize that $d_1$ and $d_2$ are the two outputs of the beam-splitter. Simple algebra gives
\\\begin{align*}
p_\t{C} &= 1 - (1-T)^{d_1^\dag d_1}- (1-T)^{d_2^\dag d_2} + (1-T)^{d_1^\dag d_1+d_2^\dag d_2} \nonumber\\
&= 1 - 2 \left(1-\frac{T}{2}\right)^{a^\dag a} + (1-T)^{a^\dag a}.
\end{align*}
Let us now assume that $T$ is much lower that the values of the photon number on the support of the state $\rho,$ that is $T \ll a^\dag a.$ In this case we can use
\\\be
(1-T)^{a^\dag a} \approx 1 - T\, a^\dag a + \frac{T^2 a^\dag a( a^\dag a -1)}{2}.
\ee
We obtain
\\\be
p_\t{C} \approx \frac{T^2}{4} a^\dag a (a^\dag a - 1) = \frac{T^2\, a^{\dag 2}a^2}{4}
\ee
and 
\\\be
p_{s1}= p_{s2} \approx \frac{T\, a^\dag a}{2}.
\ee
Combining the two previous expressions gives
\\\be
\frac{\mean{p_\t{C}}}{\mean{p_{s1}} \mean{p_{s2}}} = g^{(2)}(0).
\ee
This shows that far from the detector saturation, the way the autocorrelation is computed theoretically reproduces well what is carried out experimentally.

\section{Theory: two lasers}

\label{sec: two lasers}
\begin{figure}[t!]{}
\begin{center}
\includegraphics[width=0.5\columnwidth]{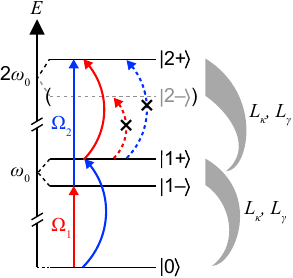}
\begin{singlespace}
\caption{\textbf{Modelled ladder states for two-laser experiment.} Scheme: laser 1 is on resonance with the $\ket{0} \leftrightarrow \ket{1-}$ transition (red arrow, detuning $\Delta_{1}=0$). Laser 2 is scanned across the $\ket{1-} \leftrightarrow \ket{2+}$ transition (blue arrow, detuning $\Delta_{2}=0$). Depicted are the driving strengths $\Omega_1$ and $\Omega_2$ of the two lasers and the decay terms of the first two rungs. The Hilbert space is truncated to $n=2$ and the state $\ket{2-}$ as well as quickly rotating off-resonant terms in the Hamiltonian are neglected. The (curved) solid arrows indicate (off-resonant) transitions taken into account in the model. The additional coincidences in the experiment (Fig.~5 of the main paper) compared to the theory can be explained by the absence of transitions to $\ket{2-}$ and $\ket{2+}$ in the model induced by laser 1 and 2, respectively (dashed curved arrows).}
\label{fig: twolas}
\end{singlespace}
\end{center}
\end{figure}

Let us now consider the case where two lasers drive the system, resulting in the driving Hamiltonian
\begin{equation}
\label{eq: H two lasers}
H_\t{L}= \left( \Omega_1 e^{\ii \omega_1 t} + \Omega_2 e^{\ii \omega_2 t} \right) a + \t{h.c.}
\end{equation}
We will assume that the two laser frequencies are almost in resonance with two transitions in the manifold of the first five levels
$\{ \ket{0}, \ket{1-}, \ket{1+}, \ket{2-}, \ket{2+}\}$, as shown in Fig.~\ref{fig: twolas}. 
The energies of the levels are given by $E_n^\pm = \pm \sqrt{n} g +\omega_\t{0} n$. The decay terms are, as before, 
\\\be
L_\kappa = \sqrt{\kappa}\, a \quad \t{and} \quad L_\gamma = \sqrt{\gamma} \,\ketbra{\t{g}}{\t{e}}.
\ee
The master equation becomes
\begin{equation}\label{eq: ME TL}
\dot \rho = -\ii [H_0+H_\t{drive}, \rho] + \sum_{\alpha=\kappa, \gamma} \left(L_\alpha \rho L_\alpha^\dag -\frac{1}{2}\{L_\alpha^\dag L_\alpha, \rho \} \right),
\end{equation}
where $H_0 = \sum_{n,\pm} E_n^\pm \proj{n\pm}$.
Note that the higher levels $\ket{3\pm}$ are not resonantly coupled to the $\ket{2\pm}$ levels, such that for small Rabi frequencies $\Omega_1$ and $\Omega_2$ one can ignore the higher levels and truncate the Hilbert space to $n=2$.

We are interested in modelling two cases with different laser frequencies. In the first case $\omega_1$ is resonant with the transition from the ground state to the lower polariton $\ket{0} \to \ket{1-}$, while $\omega_2$ is resonant with the transition $\ket{1-} \to \ket{2+}$, as shown in Fig.~\ref{fig: twolas}. That is 
\\\begin{align*}
\omega_1 &= E_1^- - 0 +\Delta_1= \omega_\t{0} - g +\Delta_1 \\
\omega_2 &= E_2^+ - E_1^- +\Delta_2= \omega_\t{0} +(1+\sqrt{2}) g +\Delta_2,
\end{align*}
where we accounted for small detunings $\Delta_1$ and $\Delta_2$ of the first and the second laser, respectively. The second case is exactly the opposite: $\omega_1$ resonant with $\ket{0} \to \ket{1+}$ and $\omega_2$ resonant with $\ket{1+} \to \ket{1-}$. The two cases are essentially equivalent, so for clarity in the following we will focus on the case depicted in Fig.~\ref{fig: twolas}.

We are interested in the regime of relatively low Rabi frequencies, in which case we can ignore the energy levels associated to $n=3,4...$ and truncate the Hilbert space to $n=2.$ Furthermore, the level $\ket{2-}$ is largely off-resonant with all possible transitions and is ignored in the simple model. Hence, we start with a four-level description of the system $\{\ket{0}, \ket{1-},\ket{1+}, \ket{2+} \}$ and the time-dependent Hamiltonian of Eq.~\eqref{eq: H two lasers}. Due to a more complicated time dependence of $H_\t{L}$ there is no rotating frame in which the Hamiltonian and the Lindblad operators are time-independent. Nevertheless, there is a good choice of a frame which, by ignoring the terms that are rotating fast, allows the physics of the system to be reproduced. This frame is
\\\begin{align*}\label{eq: RF TL}
\begin{cases}
\quad\ket{1-} \to e^{\ii \omega_1 t} \ket{1-}\\
\quad\ket{1+} \to e^{\ii \omega_2 t} \ket{1+}\\
\quad\ket{2+} \to e^{\ii (\omega_2 +\omega_1) t} \ket{2+}.
\end{cases}
\end{align*}
Considering the restriction of the operators $a$ and $\ketbra{\t{g}}{\t{e}}$, appearing in the master equation, to our subspace, we can decompose
\\\begin{align*}
\nonumber a &= \frac{1}{\sqrt{2}} \underbrace{ \ketbra{0}{1-}}_{\equiv t_{(0-)}} +\frac{1}{\sqrt{2}} \underbrace{ \ketbra{0}{1+}}_{\equiv t_{(0+)}}\\
&+\frac{\sqrt{2}+1}{2}\underbrace{\ketbra{1+}{2+}}_{\equiv t_{(+2)}} +\frac{\sqrt{2}-1}{2}\underbrace{ \ketbra{1-}{2+}}_{\equiv t_{(-2)}} \\
\ketbra{\t{g}}{\t{e}} &=\frac{1}{\sqrt{2}}(t_{(0+)} - t_{(0-)})+\frac{1}{2}(t_{(+2)} + t_{(-2)}).
\end{align*} 
In the rotating frame both these operators acquire an explicit time dependence through
\\\begin{align*}
t_{(0-)}&\to t_{(0-)}e^{-\ii \omega_1 t},\quad t_{(0+)}\to t_{(0+)}e^{-\ii \omega_2 t}, \\
t_{(-2)}&\to t_{(-2)}e^{-\ii \omega_2 t},\quad t_{(+2)}\to t_{(+2)}e^{-\ii\omega_1 t}
\end{align*}
such that the driving Hamiltonian reads
\\\begin{align*}
H_\t{L} &\approx \Omega_1 (t_{(0-)}+t_{(+2)}) + \Omega_2 (t_{(0+)}+ t_{(-2)}) +\t{h.c.},
\end{align*}
where we neglected all the quickly-rotating off-resonant terms. Taking into account the energy shifts in the rotating frame, the Hamiltonian reads
\\\be
H =
\left(
\begin{array}{cccc}
0 &\frac{\Omega_1}{\sqrt{2}}&\frac{\Omega_2}{\sqrt{2}}&\\
\frac{\Omega_1}{\sqrt{2}}& -\Delta_1& &\frac{\sqrt{2}-1}{2}\Omega_2\\
\frac{\Omega_2}{\sqrt{2}}& & -\sqrt{2}g -\Delta_2&\frac{\sqrt{2}+1}{2}\Omega_1\\
&\frac{\sqrt{2}-1}{2}\Omega_2&\frac{\sqrt{2}+1}{2}\Omega_1& -\Delta_1-\Delta_2
\end{array}
\right),
\ee
in the $\{ \ket{0}, \ket{1-},\ket{1+}, \ket{2+}\}$ basis. Note that as we assume $\Omega_1, \Omega_2, \Delta_1,\Delta_2 \ll g$ the dominant term in the Hamiltonian is $H_0 = -\sqrt{2} g \proj{1+}$.

Next, consider the decay terms $L_\kappa =\sqrt{\kappa} a$ and $L_\gamma =\sqrt{\gamma} \ketbra{\t{g}}{\t{e}}$, which also become explicitly time dependent in the rotating frame. For example, the term $L_\kappa$ in the interaction picture for $H_0$ reads
\\\begin{align*}
 L_\kappa = \sqrt{\kappa} e^{-\ii \omega_0 t} & \left( \frac{1}{\sqrt{2}}t_{(0-)}e^{\ii g t} + \frac{1}{\sqrt{2}}t_{(0+)} e^{-\ii g t} \right.\nonumber\\
 & +\frac{\sqrt{2}-1}{2} t_{(-2)} e^{-\ii(\sqrt{2}+1 )g t} \nonumber\\
 &\left. +\frac{\sqrt{2}+1}{2} t_{(+2)} e^{\ii(\sqrt{2} + 1 )g t} \right).
\end{align*}
For the master equation the global phase factor $e^{- \ii \omega_0 t}$ is irrelevant (while it plays a role for the phase of the emitted photons), because the Lindblad operators always come with the adjoint in Eq.~\eqref{eq: ME TL}. This is not the case for the relative phase terms. Yet, neglecting rapid oscillations on the time scale $1/g$ corresponds to averaging of the noise terms over short times. Formally we replace the noise term by a time average
\\\be
L_\kappa \rho L_\kappa^\dag \to \langle L_\kappa \rho L_\kappa^\dag \rangle_t,
\ee
which eliminates the coherences $\mean{e^{O(1)g t}}_t \to 0$. Effectively, this approximation leads to a new noise model with four Lindblad operators replacing $L_\kappa$
\\\begin{align*}
L_\kappa^{(0-)} &= \sqrt{\kappa}\,\frac{1}{\sqrt{2}} t_{(0-)}, \, \, \qquad L_\kappa^{(0+)} = \sqrt{\kappa} \frac{1}{\sqrt{2}} t_{(0+)}, \, \nonumber\\
L_\kappa^{(-2)} &= \sqrt{\kappa} \frac{\sqrt{2}-1}{2} t_{(-2)}, \quad L_\kappa^{(+2)} = \sqrt{\kappa} \frac{\sqrt{2}+1}{2} t_{(+2)}.
\end{align*}
Analogously, $L_\gamma$ is replaced with four noise-terms. At this point we have recovered explicitly time-independent Hamiltonian and Lindblad operators, meaning that the dynamics can be solved by the methods presented previously.

\subsection{Cavity detuning}
So far we have always assumed that the cavity mode remains at a fixed frequency $\omega_\t{C}= \omega_\t{0}$. However, the cavity can be deliberately detuned from the atomic frequency. This is particularly important for the two-laser case where the autocorrelation function $g^{(2)}(0)$ is very sensitive to the cavity detuning 
\\\be
\Delta_\text{C} =\omega_\t{C}- \omega_\t{0}.
\ee

To account for this effect we have to come back to the original Hamiltonian given in Eq.~\eqref{eq: H original}. Recalling that the free Hamiltonian is block diagonal and within each ladder rank for $n\geq 1$, spanned by $\{ \ket{\t{g},n}, \ket{\t{e},n-1}\}$, it reads
\\\be
H_\t{free}^{(n)}= \left(n \, \omega_\t{0} +(n- \frac{1}{2})\, \Delta_\text{C} \right)\Id + \frac{\Delta_\text{C}}{2} \, \sigma_z + \sqrt{n} g\, \sigma_x.
\ee
This Hamiltonian has eigenvalues
\\\begin{align}\label{eq:eigenvaluesVSdelC}
\tilde E_n^{\pm} = \left(n \, \omega_\t{0} +(n- \frac{1}{2})\, \Delta_\text{C} \right) \pm \sqrt{\Delta_\text{C}^2/4 + n g^2}
\end{align}
with corresponding eigenstates
\\\begin{align*}
 \ket{\tilde n\pm} =&\left(\frac{\Delta_\text{C}^2 }{2} + 2 n g^2 \pm \Delta_\text{C}\sqrt{\frac{\Delta_\text{C}^2}{4} + n g^2 } \right)^{-1/2}\!\!\!\!\times\nonumber \\
 &\left(\Big(\frac{\Delta_\text{C}}{2} \pm \sqrt{\frac{\Delta_\text{C}^2}{4} + n g^2 }\Big)\ket{\t{g},n} + \sqrt{n}g \ket{\t{e},n-1}\right).
\end{align*}
\begin{widetext}
In the relevant regime $\Delta_\text{C} \ll g$, this can be simplified to 
\\\begin{align*}
\tilde E_n^{\pm} \approx &\, n \omega_\t{0} \pm \sqrt{n}g + \left(n- \frac{1}{2}\right)\, \Delta_\text{C} \\
\ket{ \tilde n\pm} & \approx\frac{1}{\sqrt{2}}(\Big(1\mp \frac{\Delta_\text{C}}{4g\sqrt{n}}\Big)\ket{\t{g},n} \nonumber \\
&\pm \Big(1\pm \frac{\Delta_\text{C}}{4g\sqrt{n}}\Big)\ket{\t{e},n-1}).
\end{align*}

Following the basis change induced by the cavity emitter detuning, we also have to express the operators $a$ and $\ketbra{\t{g}}{\t{e}}$ appearing in the master equation in the basis $\{ \ket{ \tilde n\pm} \}.$ For the present purpose, we are only interested in the subspace spanned by $\{ \ket{0}, \ket{\tilde 1-}, \ket{\tilde 1-},\ket{\tilde 2+}\}$. From $\bra{\tilde n \pm} a \ket{\tilde m \pm}$ we obtain to leading order
\\\be
H=\left(
\begin{array}{cccc}
 0 & \left(\frac{1}{\sqrt{2}}-\frac{\Delta_\text{C}}{4 \sqrt{2} g}\right) \Omega _1 & \left(\frac{\Delta _\text{C}}{4 \sqrt{2} g}+\frac{1}{\sqrt{2}}\right) \Omega _2 & 0 \\
 \left(\frac{1}{\sqrt{2}}-\frac{\Delta _\text{C}}{4 \sqrt{2} g}\right) \Omega _1 & \frac{\Delta _\text{C}}{2}-\Delta _1 & 0 & \left(-\frac{\Delta _\text{C}}{8 \sqrt{2} g}-\frac{1}{2}+\frac{1}{\sqrt{2}}\right) \Omega _2 \\
 \left(\frac{\Delta _\text{C}}{4 \sqrt{2} g}+\frac{1}{\sqrt{2}}\right) \Omega _2 & 0 & -\sqrt{2} g -\Delta _2+\frac{\Delta _\text{C}}{2} & \left(\frac{\Delta _\text{C}}{8 \sqrt{2} g}+\frac{1}{2}+\frac{1}{\sqrt{2}}\right) \Omega _1 \\
 0 & \left(-\frac{\Delta _\text{C}}{8 \sqrt{2} g}-\frac{1}{2}+\frac{1}{\sqrt{2}}\right) \Omega _2 & \left(\frac{\Delta _\text{C}}{8 \sqrt{2} g}+\frac{1}{2}+\frac{1}{\sqrt{2}}\right) \Omega _1 & -\Delta _1-\Delta _2+\frac{3 \Delta _\text{C}}{2} \\
\end{array}
\right).
\ee
In addition, with the same arguments as for the non-detuned case, we obtain four noise-operators for each decay term $L_\kappa$ and $L_\gamma$
\\\begin{align*}
 L_\kappa^{(0-)} &= \sqrt{\kappa}\left(\frac{1}{\sqrt{2}} - \frac{\Delta_\text{C}}{4\sqrt{2} g}\right)\ketbra{0}{\tilde1-},
 \qquad \qquad \quad
 L_\kappa^{(0+)} = \sqrt{\kappa}\left(\frac{1}{\sqrt{2}} + \frac{\Delta_\text{C}}{4\sqrt{2} g}\right)\ketbra{0}{\tilde1+},\\
 L_\kappa^{(-2)}&=\sqrt{\kappa}\left(\frac{1}{\sqrt{2}}-\frac{1}{2} -\frac{\Delta_\text{C}}{8 \sqrt{2} g}\right)\ketbra{\tilde1-}{\tilde2+},
 \qquad 
 L_\kappa^{(+2)}=\sqrt{\kappa}\left(\frac{1}{\sqrt{2}}+\frac{1}{2} +\frac{\Delta_\text{C}}{8 \sqrt{2} g}\right)\ketbra{\tilde1+}{\tilde2+},\\
  L_\gamma^{(0-)} &= -\sqrt{\gamma}\left(\frac{1}{\sqrt{2}} + \frac{\Delta_\text{C}}{4\sqrt{2} g}\right)\ketbra{0}{\tilde1-},
  \qquad \qquad
  L_\gamma^{(0+)} = \sqrt{\gamma}\left(\frac{1}{\sqrt{2}} - \frac{\Delta_\text{C}}{4\sqrt{2} g}\right)\ketbra{0}{\tilde1+},\\
  L_\gamma^{(-2)} &= \sqrt{\gamma}\left(\frac{1}{2} - \frac{(2+\sqrt{2})\Delta_\text{C}}{16 g}\right)\ketbra{\tilde1-}{\tilde2+},
  \quad \quad
  L_\gamma^{(+2)} = \sqrt{\gamma}\left(\frac{1}{2} + \frac{(2-\sqrt{2})\Delta_\text{C}}{16 g}\right)\ketbra{\tilde1+}{\tilde2+},
\end{align*}
to leading order in $\frac{\Delta_\text{C}}{g}$.

The same line of reasoning leads to analogous expressions for the case presented in Fig.~5 of the main paper, where the first laser is close to resonance with the transition $\ket{0}\to \ket{\tilde1+}$ and the second one with $\ket{\tilde1+}\to \ket{\tilde2-}$. In this case the relevant subspace is $\{ \ket{0}, \ket{\tilde 1-}, \ket{\tilde 1+}, \ket{\tilde 2-}\}$, and we find 

\begin{equation} \label{eq: H 2L UP}
H= \left(
\begin{array}{cccc}
 0 & \left(\frac{1}{\sqrt{2}}-\frac{\Delta _\text{C}}{4 \sqrt{2} g}\right) \Omega _2 & \left(\frac{\Delta _\text{C}}{4 \sqrt{2} g}+\frac{1}{\sqrt{2}}\right) \Omega _1 & 0 \\
 \left(\frac{1}{\sqrt{2}}-\frac{\Delta _\text{C}}{4 \sqrt{2} g}\right) \Omega _2 & g \sqrt{2}-\Delta _2+\frac{\Delta _\text{C}}{2} & 0 & \left(-\frac{\Delta _\text{C}}{8 \sqrt{2} g}+\frac{1}{2}+\frac{1}{\sqrt{2}}\right) \Omega _1 \\
 \left(\frac{\Delta _\text{C}}{4 \sqrt{2} g}+\frac{1}{\sqrt{2}}\right) \Omega _1 & 0 & \frac{\Delta _\text{C}}{2}-\Delta _1 & \left(\frac{\Delta _\text{C}}{8 \sqrt{2} g}-\frac{1}{2}+\frac{1}{\sqrt{2}}\right) \Omega _2 \\
 0 & \left(-\frac{\Delta _\text{C}}{8 \sqrt{2} g}+\frac{1}{2}+\frac{1}{\sqrt{2}}\right) \Omega _1 & \left(\frac{\Delta _\text{C}}{8 \sqrt{2} g}-\frac{1}{2}+\frac{1}{\sqrt{2}}\right) \Omega _2 & -\Delta _1-\Delta _2+\frac{3 \Delta _\text{C}}{2} \\
\end{array}
\right)
\end{equation}
and 
\\\begin{align*}
 L_\kappa^{(0-)} &= \sqrt{\kappa}\left(\frac{1}{\sqrt{2}} - \frac{\Delta_\text{C}}{4\sqrt{2} g}\right)\ketbra{0}{\tilde1-},
 \qquad \qquad \quad
 L_\kappa^{(0+)} = \sqrt{\kappa}\left(\frac{1}{\sqrt{2}} + \frac{\Delta_\text{C}}{4\sqrt{2} g}\right)\ketbra{0}{\tilde1+},\\
 L_\kappa^{(-2)}&=\sqrt{\kappa}\left(\frac{1}{\sqrt{2}}+\frac{1}{2} -\frac{\Delta_\text{C}}{8 \sqrt{2} g}\right)\ketbra{\tilde1-}{\tilde2-},
 \qquad 
 L_\kappa^{(+2)}=\sqrt{\kappa}\left(\frac{1}{\sqrt{2}}-\frac{1}{2} +\frac{\Delta_\text{C}}{8 \sqrt{2} g}\right)\ketbra{\tilde1+}{\tilde2-},\\
  L_\gamma^{(0-)} &= -\sqrt{\gamma}\left(\frac{1}{\sqrt{2}} + \frac{\Delta_\text{C}}{4\sqrt{2} g}\right)\ketbra{0}{\tilde1-},
  \qquad \qquad
  L_\gamma^{(0+)} = \sqrt{\gamma}\left(\frac{1}{\sqrt{2}} - \frac{\Delta_\text{C}}{4\sqrt{2} g}\right)\ketbra{0}{\tilde1+},\\
  L_\gamma^{(-2)} &= -\sqrt{\gamma}\left(\frac{1}{2} - \frac{(2-\sqrt{2})\Delta_\text{C}}{16 g}\right)\ketbra{\tilde1-}{\tilde2-},
  \quad
  L_\gamma^{(+2)} = -\sqrt{\gamma}\left(\frac{1}{2} + \frac{(2+\sqrt{2})\Delta_\text{C}}{16 g}\right)\ketbra{\tilde1+}{\tilde2-}.
\end{align*}
\end{widetext}

\subsection{Fitting the experimental data}
Given the time-independent master equation we derived, we find the steady-state $\rho_*$ by solving
\\\be
0= - \ii [H,\rho_*] +\sum_i \left(L_i \rho_* L_i^\dag -\frac{1}{2}\{\rho_*,L_i^\dag L_i\} \right)
\ee
via the vectorization method described above. From the steady state it is easy to find the count rate
\\\be
p_S = \tr{\sum_{\alpha} L_\kappa^{\alpha} \, \rho_* L_\kappa^{\alpha\dag}},
\ee
as well as the coincidence rate
\\\be
p_\t{C} = \tr{\sum_{\alpha,\beta} L_\kappa^{\beta} \, L_\kappa^\alpha\, \rho_* L_\kappa^{\alpha \dag } L_\kappa^{\beta \dag}},
\ee
where $\alpha, \beta \in \{(0-),(0+),(-2),(+2)\}$. Both rates are functions of the parameters $\kappa, \gamma, g$ as well as the Rabi frequencies $\Omega_1,\Omega_2$, and the detunings $\Delta_1, \Delta_2, \Delta_\t{C}$.

Our aim is to use our simple model to reproduce the ``$g^{(2)}$-spectroscopy"-curves observed experimentally for the upper polariton (Fig.~5 of the main paper), i.e., the signal and the autocorrelation function $g^{(2)}(0)$ as a function of detuning $\Delta_2$ measured with detection intervals $\tau_\t{int}=155$ ps (such that oscillations on a timescale faster than $1/g$ are averaged out) . The measurements were performed at laser powers $P_1=6$ nW and $ P_2 =60$ nW, and detunings $\Delta_\t{C}= 0.31$\,GHz and $\Delta_1=0.17$\,GHz. This uniquely specifies the value of $\Omega_1 = 0.05$\,GHz, since the proportionality coefficient between the laser power and the square of Rabi frequency was calibrated with the one-laser experimental data. Nevertheless, due to the achromatic nature of the system, the value of $\Omega_2$ is not known exactly. To estimate $\Omega_2,$ we fit the $\Delta_2$ dependence of the signal by including the efficiency, that is, $\eta_{\rm det} \, p_S (\Omega_2)$ where $\eta_{\rm det}$ is the overall detection efficiency. This procedure yields $\eta_{\rm det}=10\%$ and $\Omega_2= 0.45$\,GHz. Finally, the autocorrelation function predicted by the model 
\\\be
g^{(2)}(\Delta_2) = \frac{p_\t{c}}{p_\t{S}^2}
\ee
is plotted in Fig.~ 5 of the main paper. Several remarks are appropriate.\\

First, one sees that there is a match in the $\Delta_2$-position of the peak between experiment and the model. From the model it is clear that the curve reaches its maximum at 
\\\be
\Delta_2 = \frac{3}{2}\Delta_\t{C} -\Delta_1 =0.30\, \text{GHz},
\ee
where the transition $\ket{0}\to \ket{\tilde 2-}$ is resonant with the absorption of one photon from each laser (see Eq.~\eqref{eq: H 2L UP}) such that the population of level $\ket{\tilde 2-}$ as well as the coincidence probability $p_\t{C}$ are maximized. 

Secondly, one sees that the curve obtained from the model is lower than the experimental data. This arises because our simple model underestimates the coincidence counts. At low Rabi frequencies the single counts are dominated by the emission from the first rung levels $\ket{\tilde1 \pm}$ and the coincidences arise from the emission from the levels in the second rung $\ket{\tilde 2\pm}$. While our model includes both dominant terms that populate the states of the first rung, i.e.\ the transition $\ket{0}\to_{\omega_1} \ket{\tilde 1+}$ driven by the first laser and $\ket{0}\to_{\omega_2} \ket{\tilde 1-}$ driven by the off-resonant second laser, it ignores all but one mechanism (the $\ket{0}\to_{\omega_1+\omega_2}\ket{\tilde 2-}$ two-photon transition) driving the states of the second rung. This is the price we pay for the approximations leading to the time-independent master equation, and for ignoring the $\ket{\tilde 2+}$ level.